\documentclass[lettersize,journal]{IEEEtran}
\usepackage{amsmath,amsfonts, bm}
\usepackage{algorithmic}
\usepackage{algorithm}
\usepackage{array}
\usepackage[caption=false,font=footnotesize]{subfig}
\usepackage{float}
\usepackage{subcaption}
\usepackage{svg}
\usepackage{textcomp}
\usepackage{stfloats}
\usepackage{url}
\usepackage{verbatim}
\usepackage{graphicx}
\usepackage{cite}
\usepackage{xcolor, soul}
\usepackage{siunitx}

\sethlcolor{yellow!30}  % 30% yellow, 70% white

\begin{document}

% \title{Differentiable Grouped Feedback Delay Networks: Learning from measured Room Impulse Responses for spatially dynamic late reverberation rendering %in complex spaces
% }

\title{Differentiable Grouped Feedback Delay Networks for Learning Coupled Volume Acoustics
}

\author{Orchisama Das, Gloria Dal Santo, Sebastian J. Schlecht, \IEEEmembership{Senior Member, IEEE}, Vesa V\"{a}lim\"{a}ki, \IEEEmembership{Fellow, IEEE} and Zoran Cvetkovi\'c, \IEEEmembership{Fellow, IEEE}
\thanks{This work has been accepted for publication in \emph{IEEE Transactions on Audio, Speech and Language Processing}.
This is the author’s accepted manuscript. The final published version will be
available via IEEE Xplore.}
\thanks{This work was supported by the Engineering and Physical Sciences Research Council (EPSRC) under the `Challenges in Immersive Audio Technologies' grant EP/X032981/1.}
\thanks{ O. Das and Z. Cvetkovi\'c are with the Dept. of Engineering at King's College London (email: orchisama.das@kcl.ac.uk, zoran.cvetkovic@kcl.ac.uk), G. Dal Santo (email: gloria.dalsanto@aalto.fi) and V. V\"{a}lim\"{a}ki (email:vesa.valimaki@aalto.fi) are with the Acoustics Laboratory, Dept. of Information and Communications Engineering, Aalto University, 02150 Espoo, Finland, and S. J. Schlecht is with the Dept. of Multimedia and Signal Processing at the Friedrich-Alexander-Universität Erlangen-Nürnberg, Germany (email: sebastian.schlecht@fau.de)}}

\markboth{Accepted for Publication in IEEE Transactions on Audio, Speech and Language Processing, 2026.}
{Shell \MakeLowercase{\textit{et al.}}: Bare Demo of IEEEtran.cls for IEEE Journals}


\maketitle

\begin{abstract}
Rendering dynamic reverberation in a complicated acoustic space for moving sources and listeners is challenging but crucial for enhancing user immersion in extended-reality (XR) applications. Capturing spatially varying room impulse responses (RIRs) is costly and often impractical. Moreover, dynamic convolution with measured RIRs is computationally expensive with high memory demands, typically not available on wearable computing devices. Grouped Feedback Delay Networks (GFDNs), on the other hand, allow efficient rendering of coupled room acoustics. However, its parameters need to be tuned to match the reverberation profile of a coupled space. In this work, we propose the concept of Differentiable GFDNs (DiffGFDNs), which have tunable parameters that are optimised to match the late reverberation profile of a set of RIRs captured from a space that exhibits multi-slope decay. 
Once trained on a finite set of measurements, the DiffGFDN interpolates to unmeasured locations in the space.  We propose a parallel processing pipeline that has multiple DiffGFDNs with frequency-independent parameters processing each octave band. The parameters of the DiffGFDN can be updated rapidly during inferencing as sources and listeners move. We evaluate the proposed architecture against the Common Slopes (CS) model on a dataset of RIRs for three coupled rooms. The proposed architecture generates multi-slope late reverberation with low memory and computational requirements, achieving a better energy decay relief (EDR) error and slightly worse octave-band energy decay curve (EDC) errors compared to the CS model. Furthermore, DiffGFDN requires an order of magnitude fewer floating-point operations per sample than the CS renderer.
\end{abstract}


%%%%%%%%%%%%%%%%%%%%%%%%%%%%%%%%%%%%%
% \input{sections/intro}
\section{Introduction}
\label{sec:intro}

Rendering dynamic reverberation for moving sources and listeners in complex environments is a challenging problem, but essential for immersive extended reality (XR) applications \cite{neidhardt2022perceptual, Potter_Cvetkovic_DeSena_2022}. In simple cuboid rooms with uniform absorption, late reverberation is typically approximated by an isotropic and diffuse sound field \cite{kuttruff2009roomacou, hodgson1996diffuse}, and can be well-modeled using filtered noise tails or tuned feedback delay networks (FDNs) \cite{schneiderwind2023effects, prawda2019improved}. However, in coupled spaces with different reverberation characteristics and rooms with non-uniform absorption, late reverberation becomes inhomogeneous and anisotropic, giving rise to effects such as multi-slope decay \cite{alary2021perceptual, xiang2009investigation}. Acoustic renderers in Augmented Reality (AR) applications need to model these spatially varying dynamics from limited available data, whilst being efficient and lightweight \cite{yang2022audio}.

% Coupled room reverberators
The FDN is a real-time, low-complexity artificial reverberator \cite{jot1991digital}, composed of parallel delay lines interconnected through a unitary feedback matrix \cite{gerzon1976unitary}, with additional input/output gains and absorption filters. To model multi-slope decay in such environments, we previously introduced the Grouped Feedback Delay Network (GFDN) \cite{das2020delay, das2021grouped}, which extends the FDN by grouping delay lines and associating them with distinct decay characteristics. In \cite{das2023grouped}, we further designed paraunitary feedback matrices to emulate diffraction at coupling apertures. While compact in parameterisation, the GFDN is still difficult to tune manually for measured spaces, especially for dynamic scenarios. 

The Common Slopes (CS) model \cite{Gotz_Schlecht_Pulkki_2023} offers a data-driven alternative for modelling late reverberation in coupled spaces. It represents multi-slope decays using a small set of shared decay rates and spatially varying amplitudes, enabling compact RIR modeling. During real-time rendering, position-dependent equalisers adjust the amplitudes, while static modal filters reproduce the decays \cite{Gotz_Kerimovs_Schlecht_Pulkki}. Although effective, this approach relies on computationally expensive modal reverberators, requiring hundreds of modes per band, and cannot extend to unmeasured positions. We recently proposed interpolating CS amplitudes to unseen positions using multilayer perceptrons (MLPs) \cite{Das_Cvetkovic_2025}, synthesising the late tail with filtered noise shaped by the CS parameters. However, rendering still requires time-varying convolution with the synthesised tail. In contrast, the method proposed in this paper also learns its parameters from discrete RIR measurements but eliminates convolution by replacing it with a recursive delay network.

Scerbo et al. leverage the CS model to simplify multi-slope reverberation rendering in a virtual acoustic space using FDNs 
 \cite{scerbo2025modeling, scerbo2025efficient}. The method uses acoustic radiance transfer (ART) to derive modal representations in an FDN (ART-FDN) \cite{bai2015late}, where the common decay rates are encoded in real poles, and listener/source effects are modelled via dynamic gains. Smaller FDNs with fewer delay lines and different weightings are used to render late reverberation in real-time for multiple dynamic sources and listeners. Similarly, the Coupled Volume Scattering Delay Network \cite{atalay2022scattering} targets coupled shoebox-shaped spaces. In \cite{kirsch2023computationally}, the authors propose an FDN-based, geometrically inspired reverberator for rendering coupled volume acoustics. Each sub-room in the coupled space is represented by an independent FDN, and the overall reverberation is obtained by summing their outputs, similar to the GFDN. Coupling between the volumes is modelled through weighting coefficients applied to each FDN as well as cross-mixing between their outputs. Directional late reverberation is captured by placing spatially warped virtual reverberation sources around the listener. These approaches are promising for virtual environments but not directly applicable to real measured spaces.

Recent advances in differentiable FDNs have shown that network parameters can be learned directly from measured RIRs using gradient descent. Approaches include perceptually-informed loss functions \cite{dal2023differentiable, santo2024optimisation, mezza2024data}, end-to-end training for target reverberation characteristics \cite{lee2022differentiable}, and multichannel MIMO architectures trained on spatial datasets \cite{giampiccolo2024differentiable}. However, most of these models are agnostic to spatial coordinates, limiting their ability to interpolate to unseen positions. Recently proposed differentiable Scattering Delay Networks \cite{mezza2025differentiable} can predict RIRs at unseen positions, however, they still require approximate knowledge of the room geometry, work only for shoebox rooms and cannot yet model multi-slope late reverberation observed in coupled spaces.

In this paper, we propose the Differentiable Grouped Feedback Delay Network (DiffGFDN), which learns to parameterise late reverberation with multi-slope decay directly from spatially distributed RIRs. Our method uses an MLP to map spatial coordinates of the source and receiver to corresponding source-receiver filter parameters -- analogous to the common slope amplitudes in the CS model. The delay line lengths and absorption filters, which are derived from the common decay times, are fixed. The remaining GFDN parameters, including the feedback matrix and input-output gains, are shared across space and learned globally. This allows the DiffGFDN to interpolate RIRs at unmeasured positions and adapt to listener motion in real time.

Unlike conventional FDNs or multiple-input-multiple-output (MIMO) differentiable architectures, the DiffGFDN leverages spatial input features for interpolation, reducing the need to store large RIR datasets or perform time-domain convolution during rendering. The architecture is well-suited for resource-constrained XR applications, such as AR glasses, where real-time, spatially-varying late reverberation is needed with minimal computational overhead. We demonstrate that the DiffGFDN produces RIRs with comparable energy decay properties to the CS model, while offering improved efficiency and spatial generalization.

The rest of this paper is organised as follows. In Sec.~\ref{sec:proposed_architecture}, we explain the architecture and parameters of the GFDN. In Sec.~\ref{sec:common_slope_analogies}, we derive the RIRs synthesised by the GFDN, and compare those with those of the CS model. In Sec.~\ref{sec:learning}, we discuss the configuration of the learnable parameters of the DiffGFDN and the loss functions minimised to optimise them. In Sec.~\ref{ssec:subband_parallel_gfdns}, we propose a band-wise processing architecture with multiple frequency-independent DiffGFDNs, and derive its computational and memory requirements. In Sec.~\ref{sec:eval}, we evaluate the proposed subband architecture on a three-coupled room dataset with an ablation study, and compare the energy decay curve (EDC) and energy decay relief (EDR) errors with those of the CS model for different amounts of training data. In Sec.~\ref{sec:discuss}, we discuss the benefits and limitations of the proposed method for acoustic rendering in AR. Sec.~\ref{sec:conclusion} concludes the paper.
\section{Architecture}
\label{sec:proposed_architecture}

\begin{figure*}[ht]
\centering
    \includegraphics[trim=-20 0 -20 0, clip, width=\textwidth]{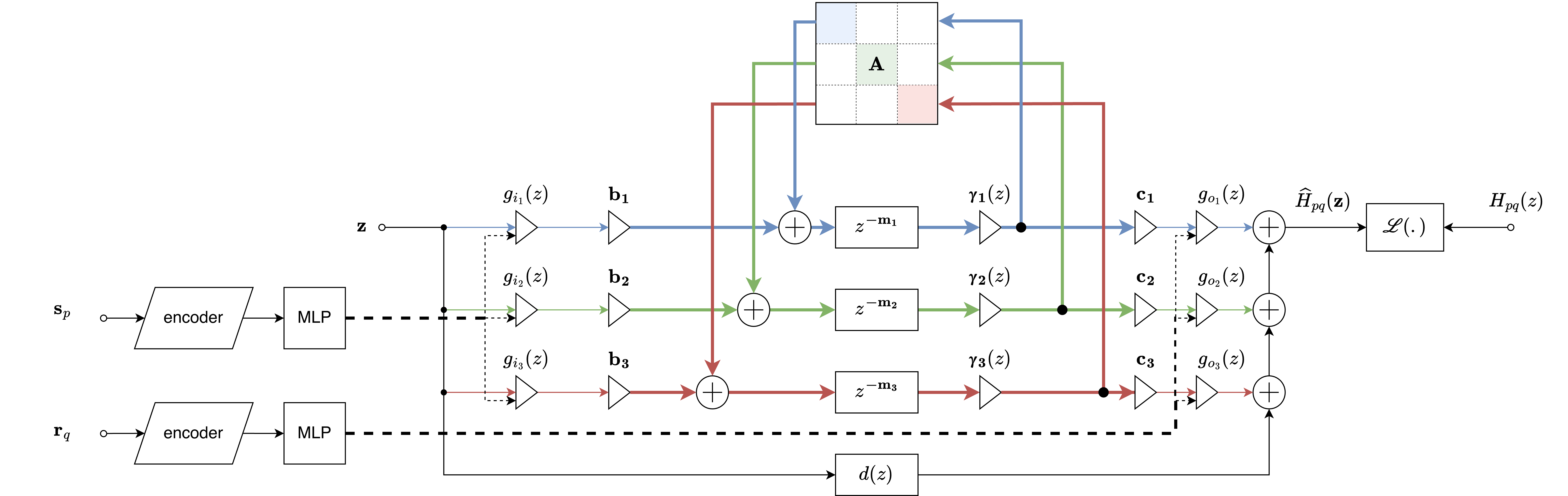}
    \caption{Differentiable Grouped FDN architecture for $G = 3$ groups. The bold lines represent multichannel signals, and the different colours represent different groups,  $\mathbf{s}_p$ and $\mathbf{r}_q$ represent source and receiver positions, respectively, and $H_{pq}(z)$ and  $\hat{H}_{pq}(z)$ are the reference and predicted transfer functions, respectively, for the position pair ($\mathbf{s}_p, \mathbf{r}_q$).}
    \label{fig:gfdn_architecture}
\end{figure*}

% A standard feedback delay network consists of $N$ delay lines of length $m_i$~samples each, $i = 1,2,\ldots,N$, with its associated absorption filter, $\gamma_i(z)$, connected through an $N \times N$ feedback matrix, $\bm{A}$. 
A standard FDN \cite{jot1991digital} consists of $N$ delay lines  of different lengths $m_i,~i=1, 2, \dots, N$ samples. Each delay line is associated with an absorption filter, $\gamma_i(z)$, connected in series, and its output is fed back to all delay inputs via an $N \times N$ feedback matrix, $\bm{A}$.
The absorption filter, specified by the reverberation time, $T_{60}$, models the frequency-dependent energy decay in the room caused by multiple reflections from the room's absorptive materials. 

In the GFDN, which we propose in this work, as shown in Fig.~\ref{fig:gfdn_architecture}, we use multiple sets of delay lines, each with a different absorption filter. Each set of delay lines with its unique reverberation time is referred to as a group. A group can represent a room in a coupled space or, within a single room, walls that have different absorption properties.
%or different walls of a single room if they have different absorption properties.
The feedback matrix can be interpreted as a block matrix controlling the amount of mixing among the different groups,
%A block feedback matrix controls the mixing among the different groups, 
with a block diagonal matrix denoting the decoupled condition and a dense matrix denoting a fully coupled system. A unitary feedback matrix  ensures energy preservation in the feedback loop. 
%The input and output gains control the relative source and listener positions. 
Unlike the GFDN architecture proposed in \cite{das2021grouped}, which has only one set of input-output gains, here we have a set of source-receiver filters, one for each group, and a different set of input-output gains, one for each delay line. The former determines the frequency-dependent amplitude variation in EDCs that occurs when changing source-receiver position, while the latter determines spectral colouration. Therefore, we decouple the position-invariant input-output gains from the position-dependent source-receiver filters.
 
The different parameters of the GFDN are listed below.
\begin{itemize}
\item $G \in \mathbb{Z}^+$ - the number of groups in the network.
\item $N \in \mathbb{Z}^+$ - the total number of delay lines in the network, here assumed to be a multiple of $G$ such that all groups have the same number of delay lines.
\item $N' = N / G \in \mathbb{Z}^+$ - the number of delay lines in each group.
\item $\bm{A} \in \mathbb{R}^{N \times N}$ - the feedback matrix.
\item $\bm{b}, \bm{c}  \in \mathbb{R}^{N \times 1}$ - the scalar input and output gains.
\item $\bm{m} = [\bm{m}_1, \cdots, \bm{m}_{G}]^T\in \mathbb{Z}^{+N \times 1}$ - the length of the delay lines, where $\bm{m}_k \in \mathbb{Z}^{+N' \times 1}$ are the delay line lengths in each group.
\item $\bm{\gamma}(z) = [\gamma_1(z), \cdots, \gamma_N(z)]^T$ - the absorption filters in the delay lines.
\item $\bm{g}_i(\mathbf{x}, z) = [g_{i,1}(\mathbf{x}, z), \cdots, g_{i,G}(\mathbf{x},z)]^T$ - the source filters (or scalar gains) in each group.
\item $\bm{g}_o(\mathbf{x}, z) \in [g_{o,1}(\mathbf{x}, z), \cdots, g_{o, G}(\mathbf{x}, z)]^T$ - the receiver filters (or scalar gains) in each group.
\item $d(z)$ - the direct and early reflection filter which models the frequency response of the direct path and early reflections.
% \item $H_{pq}(z)$ is the reference RIR at source-receiver position $(\mathbf{s}_p, \mathbf{r}_q)$.
\end{itemize}
% Unlike the GFDN architecture proposed in \cite{das2021grouped}, which has only one set of input-output gains, here we have one set of source-receiver filters (one for each group), and a different set of input-output scalars (one for each delay line). The former determines the variation in EDCs with changing source-receiver position, while the latter determines spectral colouration. Therefore, we decouple the position-invariant input-output gains from the position-dependent source-receiver filters.
% The parameters of the Differentiable GFDN are trained to match the late reverberation profile of a set of RIRs from a measured coupled space by minimising a loss function, $\mathcal{L}(.)$.
Here, $\mathbf{x} = [x, y, z]^T$ is a vector of the 3D Cartesian coordinates. 
The transfer function of the proposed GFDN structure is
    \begin{equation}
    \label{eq:gfdn_transfer_function}
    \begin{aligned}
        \hat{H}(\mathbf{x}, z) &= \bm{c}^T(\mathbf{x},z) \left(\bm{D_m}^{-1}(z)\bm{\Gamma}^{-1}(z) - \bm{A} \right)^{-1} \bm{b}(\mathbf{x}, z) + d(z), \\
        \bm{b}(\mathbf{x}, z) &= \bm{b} \odot \left(\bm{g}_i(\mathbf{x}, z) \otimes \bm{1}\right), \\ \bm{c}(\mathbf{x}, z) &= \bm{c} \odot\left (\bm{g}_o(\mathbf{x}, z) \otimes \bm{1}\right),
    \end{aligned}
    \end{equation}
where $\bm{D_m}(z) = \text{diag}[z^{-m_1}, \ldots, z^{-m_N}]$ is the diagonal matrix of delays, $\bm{\Gamma}(z) = \text{diag}[\gamma_1(z), \ldots, \gamma_N (z)]$ is the diagonal matrix of absorption filters, $\bm{1} \in \mathbb{R}^{N' \times 1}$ is a column-vector of ones, $\odot$ is the Hadamard product, and $\otimes$ is the Kronecker product. 
% \gloria{The input and output filters are computed as
%     \begin{equation}
%     \label{eq:gfdn_inout_filters}
%     \begin{aligned}
%         \bm{b}(z) &= \bm{b} \odot \left(\bm{g}_i(\mathbf{x}, z) \otimes \bm{1}\right),\\
%         \bm{c}(z) &= \bm{c} \odot\left (\bm{g}_o(\mathbf{x}, z) \otimes \bm{1}\right),
%     \end{aligned}
%     \end{equation}
% where $\bm{1}$ is a column-vector of ones $\in \mathbb{R}^{N' \times 1}$. }
The GFDN parameters fall into two categories: {\sl position-invariant parameters}, which remain constant with changing source/listener positions, and {\sl position-dependent parameters} that are updated with changing source-listener positions.
%%%%%%%%%%%%%%%%%%%%%%%%%%%%%%%%%%%%%%%%%%%%%%%%%%%%%%%%%%%%%%%%%%%%%%%%%%%%%%%%%%%%%%%%%%%%%%
\begin{itemize}

\item{Position-invariant parameters:}

\begin{itemize}
\item The delay line lengths, $\bm{m}$, which set the order of the system.
%poles, 
% and the echo density profile of the synthesised RIR.

\item The absorption filters, $\bm{\gamma}(z)$, which model the frequency-dependent energy decay of an RIR. 
% The common decay times can be found by the method suggested in \cite{Gotz_Schlecht_Pulkki_2023}, where the individual RIRs' reverberation times in octave bands are first found using a neural network \cite{Götz_Falcón_Pérez_Schlecht_Pulkki_2022}, and then clustered into $G$ groups. 
The absorption gains for the $k^\text{th}$ group and $b^\text{th}$ frequency-band, are obtained from the common decay times for the $k^\text{th}$ group and the $b^\text{th}$ subband, $T_{60_{k, b}}$ \cite{Gotz_Schlecht_Pulkki_2023}
%, and the delay line lengths for the $k^\text{th}$ group, $\bm{m_k}$ 
as,
\begin{equation}
\begin{aligned}
\label{eq:decay_time_to_gains}
\bm{\gamma}_{{k, b}_\text{dB}} &= -\frac{60}{f_s T_{60_{k, b}}} \bm{m}_k, \\
\bm{\gamma}_{k,b} &= \exp \left(-\frac{6.91}{f_s T_{60_{k, b}}} \bm{m}_k \right), 
\end{aligned}
\end{equation}
where $\bm{m}_k$ indicates the delay line lengths for the $k^\text{th}$ group.
The absorption filters, $\bm{\gamma}(z)$, can then be obtained from the band-wise gains by fitting IIR filters \cite{schlecht2017accurate, harma2000frequency}.

\item The input-output scalar gains, $\bm{b, c} \in \mathbb{R}^{N \times 1}$, which determine the spectral colouration of the network.
\item The feedback matrix, $\bm{A}$, which determines the spectral colouration and echo density. It has a unique structure given by \cite{das2021grouped},

\small
\begin{equation}
\begin{aligned}
\label{eq:feedback_matrix_structure}
\bm{A} &=
\begin{bmatrix}
\Phi_{1,1} \mathbf{M}_1^2  & \cdots & \Phi_{1,G} \mathbf{M}_1 \mathbf{M}_{G} \\ 
\Phi_{2,1} \mathbf{M}_2 \mathbf{M}_1  & \cdots & \Phi_{2, G} \mathbf{M}_2 \mathbf{M}_{G}\\
\vdots & \ddots & \vdots \\
\Phi_{G, 1} \mathbf{M}_{G} \mathbf{M}_1  &  \cdots & 
 \Phi_{G, G} \mathbf{M}^2_{G}
\end{bmatrix}, \\
 &\Phi \Phi^H = \mathbf{I}, \quad \mathbf{M}_k^H \mathbf{M}_k = \mathbf{I},
\end{aligned}
\end{equation}
\normalsize where $\mathbf{M}_k \in \mathbb{R}^{N' \times N'}$ is the unitary mixing matrix for each group, and $\Phi \in \mathbb{R}^{G \times G}$ is the unitary coupling matrix. 
\end{itemize}

\item{Position-dependent parameters:}
\begin{itemize}
\item The source filter, $\bm{g}_i(\mathbf{x}, z)$, that change according to the source position.
\item The receiver filter, $\bm{g}_o(\mathbf{x}, z)$, that change according to the listener position.
\end{itemize}
\end{itemize}

\section{Relationship with the Common Slopes Model}
\label{sec:common_slope_analogies}
In this section, we will show how the RIRs synthesised by the GFDN relate to those synthesised by the CS model \cite{Gotz_Kerimovs_Schlecht_Pulkki}. For this, we will leverage modal decomposition. Note that throughout the rest of the paper, we use the symbol $n$ to denote discrete-time steps.
%We will analyse the transfer function of the GFDN, derive its modal representation, and show that it is analogous to the modal representation of the CS model.

%%%%%%%%%%%%%%%%%%%%%%%%%%%%%%%%%%%%%%%%%%%%%%%%%%%%%%%%%%%%%%%%%%%%%%%%%%%%%%%%%%%%%%%%%

\subsection{RIRs Synthesised by the Common Slopes Model}
\label{ssec:rir_synth_cs}
The CS model proposes to model the EDC in each frequency band, as a sum of $\kappa$ exponentials, each with a different decay rate, weighted by position-dependent amplitudes,
\begin{equation}
\label{eq:common_slopes_edc}
\begin{aligned}
\mathbf{d}_b(\mathbf{x}, n) = N_{0, b} \Psi_0(n) + \sum_{k=1}^\kappa A_{k, b}(\mathbf{x}) [\Psi_{k, b}(n) - \Psi_{k, b}(L)], \\
\Psi_{k, b}(n) = 
\begin{cases}
L - n & \text{if } k = 0 \\
\exp\left(-\frac{13.8 n}{T_{60_{k, b}} f_s} \right) & \text{o.w.}
\end{cases}
\end{aligned}
\end{equation}
where $\mathbf{d}_b(\mathbf{x}, n)$ is a representation of the EDC in the $b^\text{th}$ frequency band which is a function of source-receiver positions in space, $\mathbf{x}$, and discrete time, $n$, $A_{k, b}(\mathbf{x})$ are the amplitudes for the $k^\text{th}$ slope in the $b^\text{th}$ frequency-band, which is a function of space only,  $\Psi_{k, b}(n)$ are decay kernels 
%for the $k^\text{th}$ slope in the $b^\text{th}$ frequency-band, which are 
determined by the common decay rates $T_{60_{k, b}}$,  and $N_{0,b}$ is the amplitude of the noise term in the $b^\text{th}$ frequency band, which is neglected in the following.

The final RIR, $h(\mathbf{x}, n)$, can be synthesised by shaping modes in each subband by the appropriate decay times, and applying an equalisation derived from the subband amplitudes, as suggested by Götz et al.~\cite{Gotz_Kerimovs_Schlecht_Pulkki}: 
\begin{equation}
\label{eq:rir_from_common_slopes}
\begin{aligned}
h(\mathbf{x}, n) &= \sum_{b=1}^B h_b(\mathbf{x}, n), \quad \text{where} \\
h_b(\mathbf{x}, n) &= \sum_{k=1}^\kappa   \sqrt{A_{k, b}(\mathbf{x}) \Psi_{k, b}(n)} \sum_{m=1}^{M_b} e^{j (2\pi f_{m, b} n + \varphi_{m, b})}.
\end{aligned}
\end{equation}
To get the reverberated output in real-time, a modal reverberator generates $M_b$ modes in each band, with a centre frequency of $f_b$ Hz. In total, there are $B$ frequency-bands. The mode frequencies, $f_{m,b}$, are logarithmically distributed around the centre frequency, such that,  $\frac{f_b}{\sqrt{2}} < f_{m, b} < \sqrt{2}f_b$. The mode phases, $\varphi_{m,b}$, are randomly generated between $[0, 2\pi]$~radians.

The expression for the synthesised RIR \eqref{eq:rir_from_common_slopes} can therefore be approximated in terms of convolutions between the amplitude filters and $\kappa$  modulated decay kernels representing individual EDC slopes as
\begin{equation}
\label{eq:rir_from_common_slopes_filter_form}
\begin{aligned}
h(\mathbf{x}, n) &\approx \sum_{k=1}^\kappa  \alpha_k(\mathbf{x}, n) * \Phi_k (n),
%h(\mathbf{x}, t) &= \sum_{k=1}^\kappa  \alpha_k(\mathbf{x}, t) * \sum_{b=1}^{B} \sqrt{\Psi_{k, b}(n)} \sum_{m=1}^{M_b} e^{j (2\pi f_{m,b} t + \varphi_{m,b})}, \\
% &= \sum_{k=1}^\kappa  \alpha_k(\mathbf{x}, t) * \Phi_k (n), \\
%\Phi_k (n)&=\sum_{b=1}^{B} \sqrt{\Psi_{k, b}(n)} \sum_{m=1}^{M_b} e^{j (2\pi f_{m,b} t + \varphi_{m,b})},
\end{aligned}
\end{equation}
where $\Phi_k(n)$ are the modes modulated by the $k^\text{th}$ decay kernel, 
$$\Phi_k (n) =\sum_{b=1}^{B} \sqrt{\Psi_{k, b}(n)} \sum_{m=1}^{M_b} e^{j (2\pi f_{m,b} n + \varphi_{m,b})},$$
and
$\alpha_k(\mathbf{x}, n)$ are the impulse responses of filters,
whose frequency response is specified by
$$
\mathcal{A}_k(\mathbf{x},f) = \sqrt{A_{k, b}(\mathbf{x})} \quad \text{ for } \frac{f_b}{\sqrt{2}} < f < \sqrt{2}f_b.
$$
% $$|\mathcal{A}_k(\mathbf{x},f)| = \begin{cases} \sqrt{A_{k, b}(\mathbf{x})}, \quad \text{if}\  |f| = f_b\\
% 0  \quad \text{o.w.} \end{cases}$$ 
The LHS in (\ref{eq:rir_from_common_slopes_filter_form}) would match the RHS exactly if $\alpha_k(\mathbf{x}, n) = \alpha_k(\mathbf{x}, 0)$. In real-time rendering applications, the filters, $\mathcal{A}_k(\mathbf{x},f)$,  can be implemented by fitting graphic equalisers to  $\sqrt{A_{k, b}(\mathbf{x})}$ and dynamically changing them as the source / receiver move \cite{Gotz_Kerimovs_Schlecht_Pulkki}. However, without extracting the DC component of the filters, we will not get exact equivalence.
% The resulting filters will not be narrow-band, hence the impulse response of the CS model will only be an approximation of (\ref{eq:rir_from_common_slopes}). 

% Finally this leads to
% \begin{equation}
% \begin{aligned}
% h(\mathbf{x},t) &= \sum_{k=1}^\kappa  \alpha_k(\mathbf{x}, t) * \Phi_k (n), \\
% \Phi_k (n)&=\sum_{b=1}^{B} \sqrt{\Psi_{k, b}(n)} \sum_{m=1}^{M_b} e^{j (2\pi f_{m,b} t + \varphi_{m,b})},
% \end{aligned}
% \end{equation}
% where $\Phi_k(n)$ are the modes modulated by the $k^\text{th}$ decay kernel.

%%%%%%%%%%%%%%%%%%%%%%%%%%%%%%%%%%%%%%%%%%%%%%%%%%%%%%%%%%%%%%%%%%%%%%%%%%%%%%%%%%%%%%%%%%%%%%%%%%%5
\subsection{Structure of the Feedback Matrix}
\label{ssec:structure_feedback_matrix}
A feedback matrix with no coupling between groups allows deriving close relationships between the proposed architecture and the CS model.
% We started with a no-coupling initialisation (diagonal coupling matrix), and post-training, the learned coupling matrix remained diagonal. When we tried to enforce coupling amongst the groups 
%by increasing the learning rate of the coupling angle, 
% the EDC fit was poor. 

For two groups, the GFDN has a block diagonal (decoupled) feedback matrix and its poles
% has the transfer function,
% \begin{equation}
% \begin{aligned}
%     % \hat{H}(z) &= \begin{bmatrix} \bm{c_1}^T & \bm{c_2}^T \end{bmatrix} \left( \begin{bmatrix} \bm{D_{m_1}}(z^{-1})\bm{\Gamma_1}^{-1} & 0 \\ 0 &  \bm{D_{m_2}}(z^{-1})\bm{\Gamma_2}^{-1}\end{bmatrix} - \begin{bmatrix} \mathbf{M}_1 & 0 \\
%     % 0 & \mathbf{M}_2\end{bmatrix}\right)^{-1} 
%     % \begin{bmatrix} \bm{b_1} \\ \bm{b}_2 \end{bmatrix} \\
%     \hat{H}(z) &= \sum_{i=1, 2} \bm{c}_i^T \left(\bm{D_{m_i}}(z^{-1}) \bm{\Gamma}_i^{-1} - \mathbf{M}_i \right)^{-1} \bm{b}_i\\
%     &= \frac{q_1(z)}{p_1(z)} + \frac{q_2(z)}{p_2(z)} = \frac{q_1(z)p_2(z) + q_2(z)p_1(z)}{p_1(z) \ p_2(z)},
% \end{aligned}
% \end{equation}
% where $q_i(z)$ is the residue polynomial for group $i$, which is a function of $\bm{m}_i, \bm{\Gamma}_i,\mathbf{M}_i, \bm{b}_i, \bm{c}_i$, and $p_i(z)$ is the pole polynomial for group $i$, which is a function of $\bm{m}_i, \bm{\Gamma}_i, \mathbf{M}_i$. We see that the poles of the GFDN 
are a union of the sets of poles of the two FDNs, and the residues are modified. Since the decay rates of the poles determine the decay rate of the %network
system, in the modal decomposition we expect to see two clusters of modes having two separate $T_{60}$s.
%in the modal decomposition. 
This matches the hypothesis of the CS model, which also predicts two distinct modal decay times. The zeros, on the other hand, influence the position-dependent amplitudes of the two slopes. Therefore, by altering the zeros of the GFDN (via the source-receiver filters), we can modify the position-dependent amplitudes of the CS-predicted EDCs. 

On the other hand, if we had coupling between the two groups (non-diagonal feedback matrix), the modes may no longer be separated into two groups with two distinct decay times, thereby violating the CS model. Such an architecture would introduce new modes with different frequencies and decay rates \cite{Das_Canfield-Dafilou_Abel_2019}. In reality, this is observed in coupled rooms with large apertures, especially in the low-frequency regime where new standing waves are created due to the interaction of the modes with the aperture \cite{meissner2009computer}. Therefore, the CS model, and hence, the block diagonal coupling matrix in GFDN, fails to model modal behaviour at lower frequencies.

We now derive the transfer function of the GFDN for a block-diagonal feedback matrix with no coupling. Let the input-output gains for the $k^\text{th}$ FDN in the GFDN be denoted as $\bm{b}_k,\ \bm{c}_k \in \mathbb{R}^{N' \times 1}$ and let the feedback matrix be $\mathbf{M}_k \in \mathbb{R}^{N' \times N'}$. The delay line lengths are denoted by $\bm{m}_k \in \mathbb{Z}^{+N' \times 1}$ and the diagonal matrix formed from the delays is $\bm{D_{m_k}}(z) = \text{diag} \left(z^{-\bm{m_k}} \right)$. The absorption matrix is $\bm{\Gamma}_k(z) = \text{diag}(\gamma_k(z)^{\bm{m}_k})$. The transfer function of the GFDN with the block diagonal feedback matrix ($\bm{\Phi} = \bm{I}_{G}$) can be written as
\begin{equation}
\begin{aligned}
\label{eq:gfdn_block_diag_tf}
\hat{H}(\mathbf{x}, z) &= \sum_{k=1}^{G} \bm{c}_k^T(\mathbf{x},z) \left(\bm{D_{m_k}}^{-1}(z) \bm{\Gamma}_k^{-1}(z) - \mathbf{M}_k \right)^{-1} \bm{b}_k(\mathbf{x}, z) \\ &+ d(z), %\\
%\bm{b}_k(\mathbf{x}, z) &= \bm{b}_k g_{i,k}%(\mathbf{x}, z), \qquad  \bm{c}_k(\mathbf{x}, z) %= \bm{c}_k g_{o,k}(\mathbf{x}, z).
\end{aligned}
\end{equation}
where
$$
\bm{b}_k(\mathbf{x}, z) = \bm{b}_k g_{i,k}(\mathbf{x}, z)~,~  \bm{c}_k(\mathbf{x}, z) = \bm{c}_k g_{o,k}(\mathbf{x}, z).$$
% During training, we learn the matrices $\mathbf{W}_k$ which are used to construct $\mathbf{M}_k$ in (\ref{eq:skew_sym_matrix}), and fix the coupling matrix, $\bm{\Phi} = \bm{I}_{G}$.

With the block diagonal feedback matrix, the output of the GFDN is analogous to the synthesis method proposed in \cite{Gotz_Kerimovs_Schlecht_Pulkki}, where a modal reverberator generates many modes with the desired common decay times, and then weighs them by the position-dependent amplitudes to simulate moving sources and listeners.
In the proposed architecture, we keep the feedback matrix (and therefore, the modal decay rates) fixed, and modify the source-receiver filters to alter the residues and simulate moving sources and listeners. 
% Instead of storing the amplitudes for each source-listener position, we simply store the learned weights and biases of the MLP. Since the groups in the GFDN are decoupled, the number of operations in the network is significantly reduced.

%%%%%%%%%%%%%%%%%%%%%%%%%%%%%%%%%%%%%%%%%%%%%%%%%%%%%%%%%%%%%%%%%%%%%%%%%%%%%%%%%%%%%%%%%%%%%
\subsection{RIRs Synthesised by Decoupled GFDN}
\label{ssec:rir_synth_GFDN}
With a block-diagonal feedback matrix, we can separate the GFDN into a sum of $G$ FDNs, as in (\ref{eq:gfdn_block_diag_tf}).  With a decoupled feedback matrix, we see that transfer function of the inner loop of the GFDN (without the source-receiver filters) is given by
\begin{equation}
\begin{aligned}
\label{eq:inner_loop_modal_expansion}
P(z) &= \sum_{k=1}^{G} \bm{c}_k^T \left(\bm{D}_{\bm{m}_k}^{-1}(z) \bm{\Gamma}_k^{-1}(z) - \mathbf{M}_k \right)^{-1} \bm{b}_k = \sum_{k=1}^{G} P_k(z) \\
P_k(z) &= \sum_{m=1}^M \frac{\rho_{m,k}}{1 - \lambda_{m,k} z^{-1}}. 
\end{aligned}
\end{equation}
The corresponding impulse response is,
$$
p_k(n) = \sum_{m=1}^M |\rho_{m,k}||\lambda_{m,k}|^n e^{j (\angle \rho_{m,k} + n \angle{\lambda_{m,k}})},
$$
where $\rho_{m,k}$ are the residues and $\lambda_{m,k}$ are the poles of the $m^\text{th}$ mode of the $k^\text{th}$ group. We can further divide the modes into $B$ frequency bands, and write the decomposition as
$$P_k(z) = \sum_{b=1}^B \sum_{m=1}^{M_b} \frac{\rho_{m, k, b}}{1 - \lambda_{m, k, b}z^{-1}},$$
which has the impulse response,
\begin{equation}
\begin{aligned}
\label{eq:modal_subband}
 p_k(n) &= \sum_{b=1}^B \sum_{m=1}^{M_b} |\rho_{{m, k, b}}||\lambda_{{m, k, b}}|^n e^{j (\angle \rho_{{m, k, b}} + n \angle{\lambda_{{m, k, b}}})}.
\end{aligned}
\end{equation}
Here, $\lambda_{m, k, b}$ are the poles, and $\rho_{m, k, b}$ are the residues for the $m^\text{th}$ mode in the $k^\text{th}$ group and $b^\text{th}$ frequency band. The pole frequencies follow 
$$\frac{f_b}{\sqrt{2}}< \frac{\angle{\lambda_{{m, k, b}}} f_s}{2\pi} < \sqrt{2}f_b~.$$
In Appendix~\ref{app:pole_magnitudes}, we simplify this expression further and show that the pole magnitudes are approximately constant over a frequency band, i.e., , $|\lambda_{m, k, b}| \approx |\lambda_{k,b}| = \exp \left( \frac{-6.91}{f_sT_{60_{k, b}}}\right)$. Hence, we can write the impulse response as
\begin{equation}
\label{eq:inner_loop}
\begin{aligned}
p_k(n) &\approx \sum_{b=1}^B |\lambda_{k, b}|^n \sum_{m=1}^{M_b} |\rho_{m, k, b}| e^{j (\angle \rho_{m, k, b} + n \angle{\lambda_{m, k, b}})}.
\end{aligned}
\end{equation}
Comparing $|\lambda_{k, b}|^n \approx e^{\left(-\frac{6.91 n}{T_{60_{k, b}}f_s} \right)}$ with $\Psi_{k, b}(n)$ from (\ref{eq:common_slopes_edc}) it can be concluded that  $|\lambda_{k,b}|^n \approx \sqrt{\Psi_{k, b}(n)}$.
It follows that $p_k(n)$ is closely related to the modulated modal expansion of the CS model, $\Phi_k(n)$, as defined in (\ref{eq:rir_from_common_slopes_filter_form}).
The difference between (\ref{eq:rir_from_common_slopes_filter_form}) and (\ref{eq:inner_loop}) is that the modal residues, $|\rho_{m, k, b}|$, modulate the amplitudes of the individual frequencies, unlike (\ref{eq:rir_from_common_slopes_filter_form}) which has the same amplitude modulation at all frequencies. Now, the transfer function of the GFDN can be written as
$$\hat{H}(\mathbf{x}, z) = \sum_{k=1}^{G} g_{i,k}(\mathbf{x}, z) P_k(z) g_{o,k}(\mathbf{x}, z).$$
In the time domain, the impulse response of the GFDN is
\begin{equation}
\label{eq:rir_diff_gfdn}
\begin{aligned}
    \hat{h}(\mathbf{x}, n) &= \sum_{k=1}^{G} g_{i,k}(\mathbf{x}, n) * p_k(n) * g_{o,k}(\mathbf{x}, n), \\
    &= \sum_{k=1}^{G}  g_{i,k}(\mathbf{x}, n) * g_{o,k}(\mathbf{x}, n) \ * \\
    &\left(\sum_{b=1}^B \sqrt{\Psi_{k, b}(n}) \sum_{m=1}^{M_b} |\rho_{m, k, b}| e^{j (n \angle{\lambda_{m, k, b}} + \angle \rho_{m, k, b})} \right). 
\end{aligned}
\end{equation}
Here, $g_{i,k}(\mathbf{x},z), g_{o,k}(\mathbf{x}, z)$ are the $k^\text{th}$ elements of the position-dependent $\bm{g}_i(\mathbf{x}, z), \bm{g}_o(\mathbf{x}, z)$, and $g_{i, k}(\mathbf{x}, n), g_{o, k}(\mathbf{x}, n)$ are their corresponding impulse responses in the time domain.  
% In our proposed model, we wish to learn these filters with a neural network. We can interpolate the filters at a position where the RIR is unmeasured, using the trained network. 

% The term $g_{i_{k}}(n) *  g_{o_{k}}(n)$ is analogous to the GEQ fitted to the subband amplitudes for the kth slope, $\alpha_{k}(\mathbf{x}, t)$, in (\ref{eq:rir_from_common_slopes_filter_form}). However, in this case, each mode is scaled by the magnitude of its residue,  $|\rho_{m, k, b}|$, unlike (\ref{eq:rir_from_common_slopes_filter_form}).

It has been shown that minimising the deviation from an allpass magnitude response makes the distribution of the residue magnitudes narrower \cite{heldmann2021role, dal2023differentiable}. In fact, an allpass FDN has the narrowest distribution of residues. Thus, incorporating a spectral loss function that penalises deviation from a flat magnitude response in training the GFDN ensures that $|\tilde{\rho}_{m, k, b}| \approx |\bar{\rho_k}|$. Using this approximation, (\ref{eq:inner_loop}) reduces to,
\begin{equation}
\label{eq:inner_loop_reduced}
\begin{aligned}
p_k(n) &\approx |\bar{\rho_k}| \ \sum_{b=1}^B \sqrt{\Psi_{k, b}(n)} \sum_{m=1}^{M_b} e^{j ( n \angle{\lambda_{m, k, b} + \angle \rho_{m, k, b})}}, \\
&\approx |\bar{\rho_k}| \tilde{\Phi}_k(n).
\end{aligned}
\end{equation} 

Now, from (\ref{eq:rir_diff_gfdn}) and (\ref{eq:inner_loop_reduced}), we can write the impulse response of the GFDN as:
\begin{equation}
\label{eq:rir_diff_gfdn_final_form}
    \hat{h}(\mathbf{x}, n) = \sum_{k=1}^{G} g_{i,k}(\mathbf{x}, n) * g_{o,k}(\mathbf{x}, n) * (|\bar{\rho_k}| \tilde{\Phi}_k(n)).
\end{equation}
The term $g_{i,k}(\mathbf{x}, n) *  g_{o,k}(\mathbf{x}, n)$ is analogous to the GEQ fitted to the subband amplitudes for the $k^\text{th}$ slope. Compared to (\ref{eq:rir_from_common_slopes}), there is an additional scaling term, $|\tilde{\rho}_k|$, which is the magnitude of the mode residues for the $k^\text{th}$ group.  Note that the mode frequencies and phases are not equal to those of the CS model and therefore, $\Phi_k(n) \neq \tilde{\Phi}_k(n)$. However, they are analogous and represent wide-band signals that are statistically similar to white noise.

%%%%%%%%%%%%%%%%%%%%%%%%%%%%%%%%%%%%%%%%%%%%%%%%%

\section{Proposed Parameter Learning}
\label{sec:learning}

% In this section, we describe the learnable parameters of the network and the loss function minimised to optimise them \gloria{via stochastic gradient descent}. We refer to the GFDN with learnable parameters as the DiffGFDN. The parameters of the DiffGFDN are trained to match the late reverberation profile of a set of RIRs from a measured coupled space by minimising a differentiable loss function. 
Starting from a set of RIRs measured at different source-receiver positions, we would like to train the parameters of the GFDN, so that it can interpolate late reverberation at unseen source-receiver positions. In this section, we describe the fixed and learnable parameters of the GFDN, and the differentiable loss functions used for optimisation via stochastic gradient descent. We refer to the GFDN with learnable parameters as the DiffGFDN. The loss functions are designed to capture the energy decay profile of the late reverberation being modelled.
\subsection{Fixed Parameters}
The delay line lengths, $\bm{m}$, and the absorption filters, $\bm{\gamma}(z)$ are fixed in the network.
\begin{itemize}
\item The delay line lengths, $\bm{m}$, are typically set to be co-prime numbers to avoid 
% comb filtering effects 
degenerated patterns in the echo density, leading to temporal fluctuations, slow echo build-up, and temporal sparsity
\cite{schlecht2016feedback}.
\item  The absorption filters, $\bm{\gamma}(z)$, are derived from the common decay times in octave bands. The common decay times can be found by the method suggested in \cite{Gotz_Schlecht_Pulkki_2023}, where the individual RIRs' reverberation times in octave bands are first found using a neural network \cite{Götz_Falcón_Pérez_Schlecht_Pulkki_2022}, and then clustered into $G$ groups. The common decay times are converted to absorption gains using (\ref{eq:decay_time_to_gains}). The absorption filters are modelled with a graphic equaliser (GEQ) with 
%centre frequencies in octave bands
one band per octave, and optimised command gains.
%, using the constrained magnitude least squares method proposed in \cite{schlecht2017accurate}. 
The GEQ is designed as a cascade of biquad filters with fixed center frequencies at the octave bands.
% The GEQ consists of a cascade of biquads which are parameterised by fixed cutoff frequencies at octave bands and learnable command gains.
% where the first and the last biquads are low and high shelving filters respectively. These are parametrised by cutoff frequencies at the first and last octave bands, and learnable command gains. The other biquads are peak filters that are parameterised by cutoff frequencies (at the remaining octave bands), learnable command gains, and a fixed Q factor. 
The optimum command gains are found via a constrained linear least squares method \cite{schlecht2017accurate}, by minimising the $\ell_{2}$-norm of the deviation of the GEQ's magnitude response of the desired absorption filter gains, $$\bm{\gamma}_{k,b_\text{dB}}(\omega_\mathrm{P}) = \frac{-60 \bm{m}}{f_s T_{60}(\omega_\mathrm{P})}~,$$ where $T_{60}(\omega_\mathrm{P})$ are the reverberation times at logarithmically placed control frequencies, $\omega_\mathrm{P}$. For more information on the structure of a GEQ, please see \cite{valimaki2016all}.

\end{itemize}

%%%%%%%%%%%%%%%%%%%%%%%%%%%%%%%%%%%%%%%%%%%%%%%%%%%%%%%%%%%%%%%%%%%%%%%
\subsection{Learnable Parameters}

\begin{itemize}
\item The position-invariant input-output scalar gains, $\bm{b, c} \in \mathbb{R}^{N \times 1}$ are learned via gradient descent.

\item The position-invariant block-diagonal feedback matrix, $\bm{A}$, has the structure,

\small
\begin{equation}
\begin{aligned}
\label{eq:feedback_matrix_structure}
\bm{A} &=
\begin{bmatrix}
\mathbf{M}_1^2  &  \cdots & \bm{0} \\ 
\vdots & \ddots & \vdots \\
\bm{0}  &  \cdots & 
\mathbf{M}^2_{G}
\end{bmatrix}~.\\
\end{aligned}
\end{equation}
\normalsize The individual unitary feedback matrices, $\mathbf{M}_k$, are represented as exponentiated skew-symmetric matrices to ensure orthogonality \cite{dal2023differentiable},  
\begin{equation}
    \label{eq:skew_sym_matrix}
    \mathbf{M}_k = \exp\left(\mathbf{W}_{k,\text{Tr}} - \mathbf{W}_{k,\text{Tr}}^T \right),
\end{equation}
where $\mathbf{W}_{k, \text{Tr}}$ is the upper triangular part of a unconstrained real-valued matrix, $\mathbf{W}_k \in \mathbb{R}^{N' \times N'}$ and $\exp(.)$ denotes the matrix exponential. The matrices, $\mathbf{W}_k$, are learned via gradient descent. {Note that a position-invariant feedback matrix is needed for analogy with the CS model. It ensures the inner loop of the GFDN}, (\ref{eq:inner_loop}), {produces noise shaped by $G$ decay kernels with $G$ distinct decay profiles, which are weighted by the position-dependent source-receiver filters.}

\item The parameters of the position-dependent source-receiver filters, $\bm{g}_i(\mathbf{x}, z), \bm{g}_o(\mathbf{x}, z)$, are learned with an MLP. The inputs to the MLP are 3D Cartesian coordinates of the source/receiver positions, encoded with Fourier transformations \cite{richard2022deep} to enable the MLP to learn variations in the data at high spatial frequencies,
\begin{equation}
\label{eq:mlp_inputs}
\begin{aligned}
\eta(\mathbf{x}) &= \big[ 
\sin(\pi \ell_1 \mathbf{x}), \ldots, \sin(\pi \ell_J \mathbf{x}), \nonumber \\
& \hspace{2em} \cos(\pi \ell_1 \mathbf{x}), \ldots, \cos(\pi \ell_J \mathbf{x}) 
\big]^T \\
    % \eta(\mathbf{x}) &= \begin{bmatrix}
    % \sin(\pi \ell_1 \mathbf{x}) &
    % \cdots &
    % \sin(\pi \ell_L \mathbf{x}) \\
    % \cos(\pi \ell_1 \mathbf{x}) &
    % \cdots &
    % \cos(\pi \ell_L \mathbf{x})
    % \end{bmatrix}^T\\
    \ell_n &= f_{\textrm{min}} \left(\frac{f_{\textrm{max}}}{f_{\textrm{min}}} \right)^{\frac{n}{N_f-1}}, \quad n = [0, \ldots, J-1],
\end{aligned}
\end{equation}
where the sine and cosine functions are applied element-wise to $\mathbf{x}$, and  $N_f$ is the number of input frequencies. The output of the MLP are either scalar gains, or state variable filter (SVF) coefficients that are converted to parametric equalisers (PEQ) \cite{lee2022differentiable} that represent the source and receiver filters, $\bm{g}_i(\mathbf{x
}, z), \bm{g}_o(\mathbf{x}, z)$. More detail on the MLP output is provided in Sec.~\ref{ssec:subband_parallel_gfdns}.
% The PEQ consists of a cascade of biquad filters, each of which is parameterised by a fixed centre frequency, and learnable resonances and command gains.
\end{itemize}
%%%%%%%%%%%%%%%%%%%%%%%%%%%%%%%%%%%%%%%%%%%%%%%%%%%%%%%%%%%%%%%%%%%%%%%%%%%%%%%%%%%%%%%%%%
\subsection{Loss Functions}

Our loss functions use frequency-sampling \cite{lee2022differentiable} to evaluate the transfer function of the DiffGFDN for a source-receiver pair at $(\mathbf{s}_p, \mathbf{r}_q)$ at uniformly distributed $Q$ discrete points on half the unit circle, $z = [e^{\frac{\pi j.0}{Q}}, \cdots, e^{\frac{\pi j (Q-1)}{Q}}]$,
\begin{equation}
\label{eq:frequency_sampled_tf}
    \hat{H}_{pq}(z) = \bm{c}^T(\mathbf{r}_q,z) \left[\bm{D_m}^{-1}(z) \bm{\Gamma}^{-1}(z)- \bm{A}\right]^{-1} \bm{b}(\mathbf{s}_p, z) + d(z).
\end{equation}
Filters $\bm{b}(\mathbf{s}_p, z)$ and $\bm{c}(\mathbf{r}_q, z)$ are further decomposed as an element-wise multiplication of position-independent scalar gains, $\bm{b}, \bm{c} \in \mathbb{R}^{N \times 1}$, with position-dependent filters, $\bm{g}_i(\mathbf{s}_p, z)$, $\bm{g}_o(\mathbf{r}_q, z)$ as in (\ref{eq:gfdn_transfer_function}). 

% Since the source-receiver filters are IIR, they introduce new poles in the system with decay rates different from the common decay times. We observed through analysis that the decay rates of these new poles are much faster than the modal decay rates of the DiffGFDN, and therefore, can be ignored as they do not affect the energy decay properties of the output. Alternatively, IIR filters can be replaced with FIR filters, but the number of FIR taps needed will be much larger, making training more difficult and adding additional complexity during inferencing.
%\textcolor{red}{ZC: be more speficic, e.g. say they do not change EDC, which are the primary focus of this paper.} 

The room impulse response in the time-domain, $\hat{h}_{pq}(n)$, is found by taking an inverse DFT of the transfer function, 
% \textcolor{red}{ZC: that is inverse DFT, if you invert the z-transform at equidistant points on the unit circle, otherwise it is just the z-transform}
$\hat{H}_{pq}(z)$. To ensure adequate sampling, the number of sampling points is set to  $Q = 2^{\lceil \log_2(T_{60_{\textrm{max}}} f_s) \rceil}$, where $T_{{60}_\textrm{{max}}}$ is the maximum reverberation time of the measured room impulse responses, $f_s$ is the sampling frequency and $\lceil . \rceil$ is the ceiling operator. The time domain signal also has a length of $Q$ samples.

\subsubsection{Energy Decay Loss Functions}
To match the energy decay of a measured transfer function, $H_{pq}(z)$, at a source-receiver location, $(\mathbf{s}_p, \mathbf{r}_q)$, we minimise the normalised energy decay relief (EDR) loss between the DiffGFDN's output and the reference RIR at each location \cite{Mezza_Giampiccolo_Bernardini_2024}. The EDR characterises the energy decay of an RIR as a function of frequency and time, and is defined by the squared magnitude of the short-time Fourier transform (STFT) of the impulse response \cite{jot1992analysis}. We form an EDR loss function, $\mathcal{L}_{\text{EDR}_{pq}}$, based on the EDR of the reference transfer function, $\text{EDR}_{pq}(k, n)$, and the EDR obtained from the GFDN's transfer function, $\widehat{\text{EDR}}_{pq}(k, j)$, where $j$ is the time index and $k$ is the frequency bin of the STFT which has a total of $J$ time indices:

\small
\begin{equation}
\begin{aligned}
\text{EDR}_{pq}(k, j) &= 10 \log_{10} \left(\sum_{\tau=j}^J |H_{pq}(k, \tau) |^2 \right), \\
\widehat{\text{EDR}}_{pq}( k, j) &= 10 \log_{10} \left(\sum_{\tau=j}^J |\hat{H}_{pq}(k, \tau) |^2 \right),  \\
\mathcal{L}_{\text{EDR}_{pq}} &= \frac{ \sum_k \sum_h |\text{EDR}_{pq}(k, j) - \widehat{\text{EDR}}_{pq}(k, j)|}{\sum_k \sum_j |\text{EDR}_{pq}(k, j)|}.
\end{aligned}
\end{equation}
\normalsize

We also add a broadband energy decay curve (EDC) loss \cite{Mezza_Giampiccolo_Bernardini_2024}. The EDC loss, $\mathcal{L}_{\text{EDC}_{pq}}$, is the mean absolute difference between the desired EDC, $\text{EDC}_{pq}(n)$, obtained from the reference room impulse response, $h_{pq}(n)$,  and the final EDC, $\widehat{\text{EDC}}_{pq}$(n), obtained from the GFDN's impulse response, $\hat{h}_{pq}(n)$. The EDCs are calculated in decibels, and evaluated at the source-receiver position pair, $(\mathbf{s}_p, \mathbf{r}_q)$ as,
\begin{equation}
\begin{aligned}
\label{eq:edc_loss}
\text{EDC}_{pq}(n) =& 10\log_{10}\left(\sum_{l=n}^Q h_{pq}^2(l) \right), \\
\widehat{\text{EDC}}_{pq}(n) &= 10\log_{10}\left(\sum_{l=n}^Q \hat{h}_{pq}^2(l) \right), \\
\mathcal{L}_{\text{EDC}_{pq}} &= \frac{1}{Q} \sum_{n=1}^Q \left|\text{EDC}_{pq}(n) - \widehat{\text{EDC}}_{pq}(n)\right|.
\end{aligned}
\end{equation}
We incorporate random masking of time indices in calculating the EDC loss to introduce stochasticity. The masked indices are derived from a Bernoulli distribution with probability $0.5$. To get a scalar loss when training over batches, we average $\mathcal{L}_{\text{EDR}_{pq}}$ and $\mathcal{L}_{\text{EDC}_{pq}}$ over all source-receiver pair positions in the batch to get $\mathcal{L}_{\text{EDR}}$ and $\mathcal{L}_{\text{EDC}}.$

%%%%%%%%%%%%%%%%%%%%%%%%%%%%%%%%%%%%%%%%%%%%%%%%%%%%%%%%%%%%%%%%%%%%%%%%%%%%%%%%%%%%%%%%%%%%%%%%
\subsubsection{Colouration and Sparsity Losses}
\label{sssec:spec_sparsity_loss}
In the loss function we include two additional terms to control the overall colouration of the GFDN’s response and its echo density. This step has been shown to help reduce common artefacts associated with FDNs that use few delay lines \cite{dal2023differentiable, santo2024feedback}.
The spectral colouration loss term penalises deviation of each group in the GFDN from that of an allpass magnitude response.  
%We find the optimum parameters to minimise colouration of each group in the GFDN. 
%For the $k^\text{th}$ group in the GFDN, we add a spectral colouration loss term that penalises deviation from an allpass magnitude response,

\begin{equation}
\begin{aligned}
\label{eq:spectral_loss}
\mathcal{L}_{\text{spectral}_k} &= \frac{1}{Q} \sum_{q=0}^{Q-1} \left(|\bm{c}_k^T \left(\bm{D_{m_k}}^{-1}(\omega_q) - \mathbf{M}_k \right)^{-1} \bm{b}_k| - 1\right)^2,
\end{aligned}
\end{equation}
where $\omega_q = \frac{2\pi q}{Q}$ are discrete points on the unit circle. The absorption filters are ignored here, and we only look at the marginally stable lossless FDN prototype \cite{dal2023differentiable}. This is valid because the modal residues, with and without absorption, remain approximately the same. Hence, we can derive (\ref{eq:inner_loop_reduced}) from (\ref{eq:inner_loop_modal_expansion}) by minimising colouration of the marginally stable prototype (see Appendix~\ref{app:modal expansion}).

To ensure smooth reverberation and adequate echo density \cite{schlecht2020scattering}, 
% we add another additional a loss term that
the sparsity loss term penalises 
%the sparsity of 
each FDN's response. This term ensures that the feedback matrices, $\mathbf{M}_k$'s, are dense and is given by \cite{santo2024feedback},
\begin{align*}
\mathcal{L}_{\text{sparsity}_k} = \frac{N' \sqrt{N'} - \sum_{i,j}|\mathbf{M}_k(i,j)|}{N'\sqrt{N'} - 1}.
\end{align*}
Without this additional loss, the spectral loss optimisation yields diagonal feedback matrices, $\mathbf{M}_k$, and the network becomes a parallel combination of comb filters. 

The overall loss is, therefore, 
\begin{equation}
\begin{aligned}
    \label{eq:total_loss}
    \mathcal{L}_{\text{total}} = &\lambda_{\textrm{EDC}}\mathcal{L}_{\text{EDC}} + \lambda_{\text{EDR}}\mathcal{L}_{\text{EDR}} + \\
    &\sum_{k=1}^{G} \left(\lambda_{\text{spectral}}\mathcal{L}_{\text{spectral}_k} + \lambda_{\text{sparsity}}\mathcal{L}_{\text{sparsity}_k} \right),
\end{aligned}
\end{equation}
where $\lambda$'s are the weighting factors for the corresponding loss terms.
\section{Subband Processing with Parallel GFDNs}
\label{ssec:subband_parallel_gfdns}

\begin{figure*}
\centering
\subfloat[Parallel subband processing -- training pipeline. 
The dashed lines represent information from the dataset to the DiffGFDNs. The reference room transfer functions (RTFs) are first passed through a reconstructing full-octave filter bank. Then, each of the $B$ filtered subband RTFs is sent to the loss function calculator. Each subband DiffGFDN's transfer function is passed through the same filters as the RTFs before being sent to the loss function calculator.
]{
 \includegraphics[trim={0 10 0 10},clip,width=0.45\textwidth]{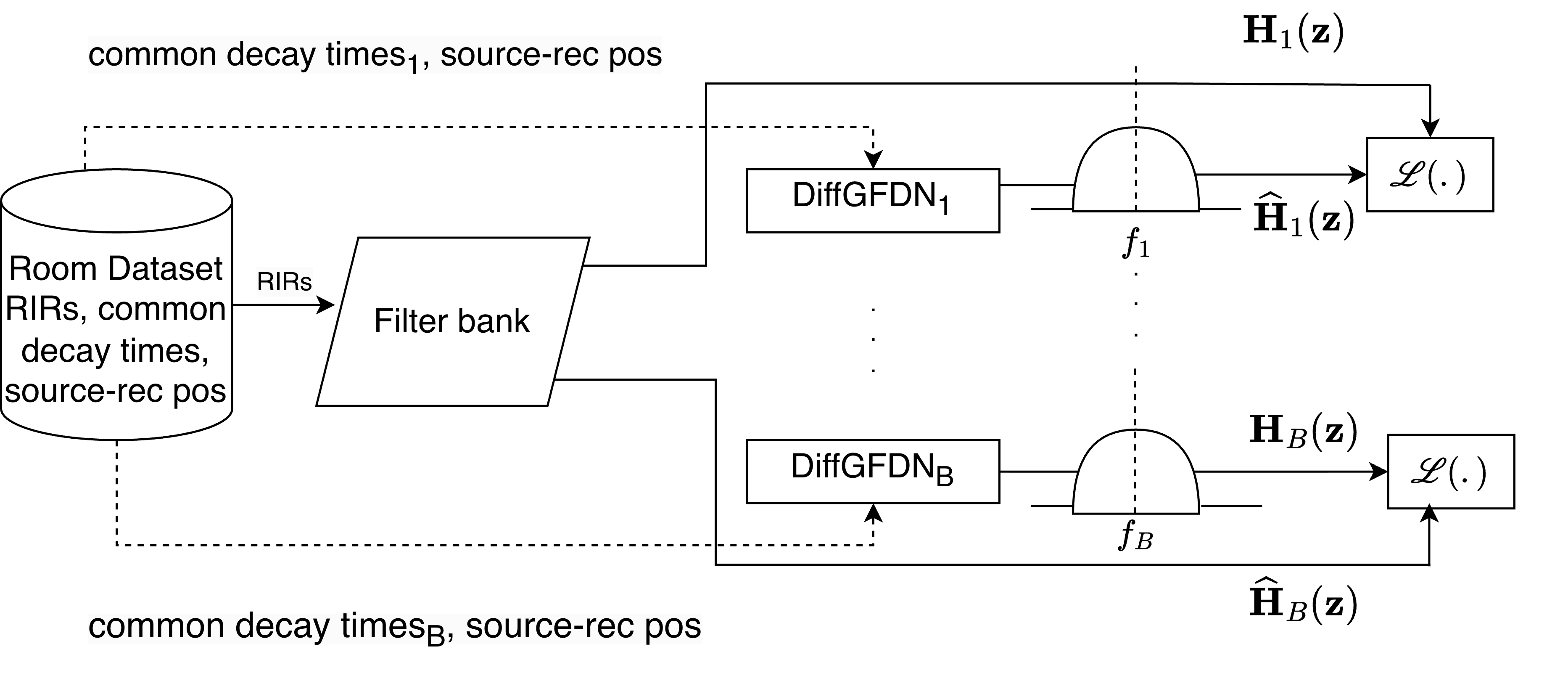}
\label{subfig:subband_training}
} \qquad
\subfloat[Parallel subband processing -- inferencing. 
The incoming signal is first passed through a filter bank and each of the $B$ subbands is processed by a trained subband DiffGFDN. Additional source-receiver information is provided to the DiffGFDNs (shown in dashed lines). The outputs of each DiffGFDN are summed to get the final reverberated output.
]{
 \includegraphics[trim={0 50 0 20}, clip, width=0.45\textwidth]{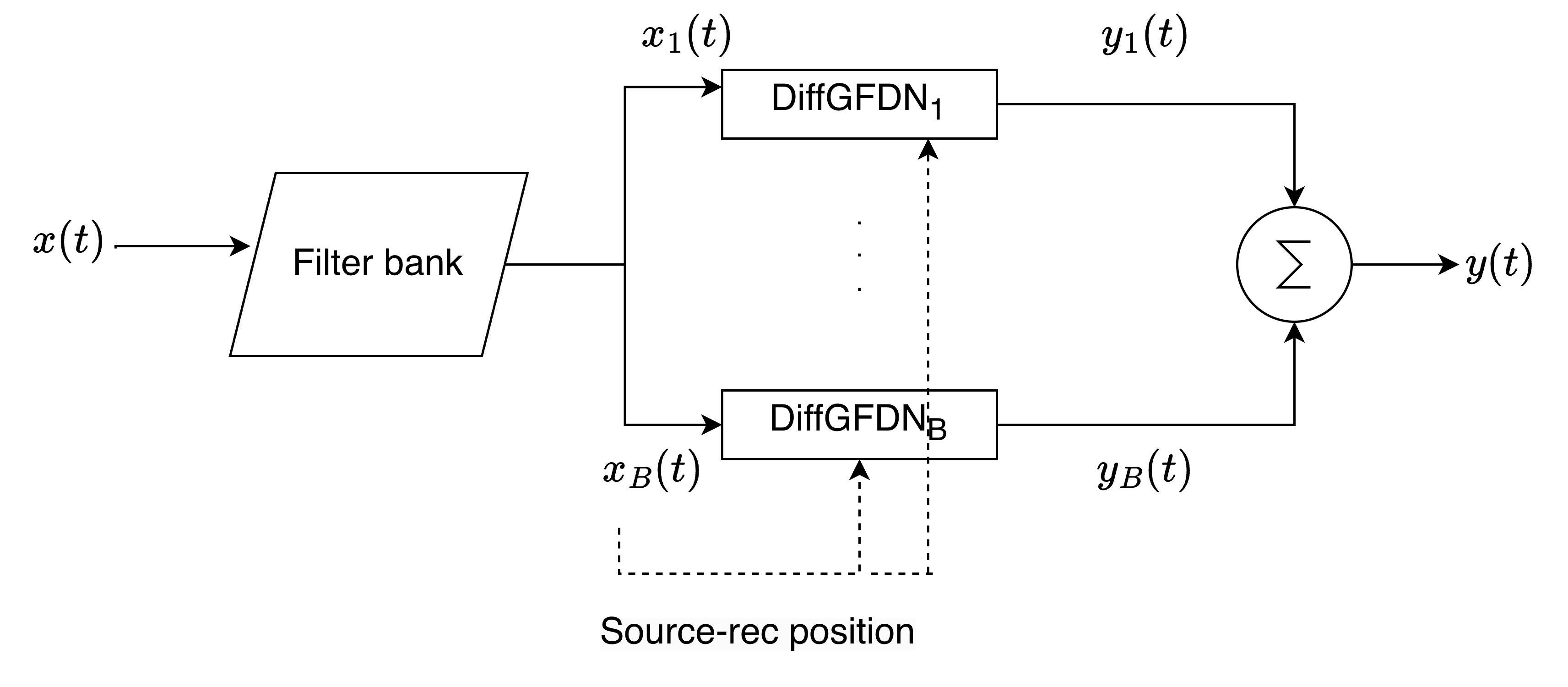}
 \label{subfig:subband_inferencing}}
\caption{Training (left) and inferencing (right) pipelines for subband processing of the input signal with a parallel network of DiffGFDNs.}
\label{fig:subband_processing}
\end{figure*}

To modulate the EDC amplitudes, we can process each frequency band with a separate DiffGFDN with source and receiver gains instead of using a single DiffGFDN with source and receiver filters. We do this because using source-receiver filters will not yield equivalence between the desired and achieved RIRs (see Sec.~\ref{ssec:rir_synth_cs}). Moreover, learning the parameters of frequency-sampled, differentiable IIR filters through deep learning poses several challenges. To avoid time-aliasing, longer responses must be evaluated, which can increase training time \cite{hayes2024review}. The estimated coefficients must be constrained to ensure stability \cite{lee2022differentiable, kuznetsov2020differentiable}, which can increase the roughness of the loss landscape, requiring smoothing techniques and fine tuning. In addition, higher-order filters require deeper MLPs to learn their parameters \cite{colonel2022direct}. Therefore, we prefer to process each frequency band separately to simplify the loss landscape and reduce overall complexity.
% the parameter space can be either overly large or too constrained 
% Moreover, learning the parameters of frequency-sampled, differentiable IIR filters using deep learning presents several challenges. Depending on the chosen filter design and activation functions, the parameter space can be either overly large or too constrained \cite{colonel2022direct, hayes2024review}. 
% Moreover, the number of biquad filters that can be cascaded is limited due to instability and numerical issues . 

In \cite{lee2022differentiable}, the authors learned the parameters of differentiable parametric equalisers (PEQs) that consist of a cascade of biquad filters parametrised as SVFs. We tried a similar parameterisation, and fixed the centre frequencies of the PEQs at octave bands, and attempted to learn the resonance and command gains of each biquad with the MLP. This resulted in large EDC errors in lower frequencies (demonstrated in Sec.~\ref{ssec:ablation}). Therefore, we propose an alternate architecture where each frequency band of a signal is processed with a different GFDN, all of whose parameters are frequency-independent.

 For $B$ frequency bands, this results in $B$ GFDNs—each operating with frequency-independent decay characteristics, determined by the common decay time in its respective band. The delay line lengths $\bm{m}_k$ and absorption gains $\bm{\gamma}_{k,b}$ are fixed. Each network learns a feedback matrix $\mathbf{M}_{k,b}$, input/output gains $\bm{b}_{k,b}$, $\bm{c}_{k,b}$, position-dependent source gains, $\bm{g}_{i,b} = [g_{{i, 1,b}}, \dots, g_{{i, G, b}}]^T \in \mathbb{R}^{G \times 1}$, and receiver gains, $\bm{g}_{o,b} = [g_{{o, 1, b}}, \dots, g_{{o, G_b}}]^T \in \mathbb{R}^{G \times 1}$, for each band.

Each MLP outputs a vector of size $G$,
% $G \times 1$, 
leading to a total of $2 \times B \times G$ outputs across all $B$ MLPs, i.e.,
\begin{equation}
\begin{aligned}
\text{MLP}_b(\eta(\mathbf{s}_p)) &\rightarrow \bm{g}_{i,b}, \qquad
\text{MLP}_b(\eta(\mathbf{r}_q)) \rightarrow \bm{g}_{o,b}.
\end{aligned}
\end{equation}

The training pipeline, shown in Fig.~\ref{subfig:subband_training}, uses a perfect-reconstruction FIR filter bank \cite{antoni2010orthogonal} to split each measured RIR into $B$ frequency bands. The outputs of the individual GFDNs are passed through the corresponding filters before computing the loss.

At inference time, the input signal is similarly decomposed into $B$ subbands using the same filter bank. Each subband is processed by its corresponding DiffGFDN in parallel, and the outputs are summed to produce the reverberated signal. This inference pipeline is illustrated in Fig.~\ref{subfig:subband_inferencing}. The final transfer function of the system, $\hat{H}(\mathbf{x}, z)$, is given by,

\vspace{-1em}
\footnotesize
\begin{equation}
\begin{aligned}
\hat{H}(\mathbf{x}, z) &= \sum_{b=1}^{B} \hat{H}_b(\mathbf{x},z) G_b(z), \\
\hat{H}_b(\mathbf{x}, z) &= \sum_{k=1}^{G} g_{{i, k, b}}(\mathbf{x}) \left[\bm{c}_{k, b}^T \left(\bm{D}_{\bm{m_k}}^{-1}(z) \bm{\Gamma}_{k, b}^{-1} - \mathbf{M}_{k, b} \right)^{-1} \bm{b}_{k, b} \right] g_{{o, k, b}}(\mathbf{x}),
\end{aligned}
\end{equation}
\normalsize 
where $G_b(z)$ is the transfer function of the $b^\text{th}$ subband filter, and $\bm{\Gamma}_{k,b} = \text{diag} \left(\bm{\gamma}_{k, b}\right)$ is the diagonal matrix of absorption gains, derived from the common decay times, $T_{60_{k,b}}$, using (\ref{eq:decay_time_to_gains}). Using (\ref{eq:inner_loop_modal_expansion}), (\ref{eq:inner_loop_reduced}) and (\ref{eq:rir_diff_gfdn_final_form}), we can write the impulse response of the parallel subband GFDNs as,
\begin{equation}
\begin{aligned}
    \hat{h} (\mathbf{x}, n) &=  \sum_{b=1}^B \hat{h}_b(\mathbf{x}, n), \\
    \hat{h}_b(\mathbf{x}, n) &\approx 
    % \sum_{k=1}^{G}  |\bar{\rho}_{k}| g_{i_{k, b}}(\mathbf{x}) \ g_{o_{k, b}}(\mathbf{x})  \sqrt{\Psi_{k, b}(n)} \ \sum_{m=1}^{M_b}  e^{j (\angle{\rho_{k, m, b}} + t \angle{\lambda_{k, m, b}})}, \\
    \sum_{k=1}^{G}  |\bar{\rho}_{k}| g_{{i,k, b}}(\mathbf{x}) \ g_{{o,k, b}}(\mathbf{x}) \tilde{\Phi}_{k, b}(n).
\end{aligned}
\end{equation}
\normalsize
% This expression is a scaled version of (\ref{eq:rir_from_common_slopes}), with $g_{i_{k, b}}(\mathbf{x}) \ g_{o_{k, b}}(\mathbf{x}) = \sqrt{\bm{A}_{k, b}(\mathbf{x})}$ and a modulation factor of $|\rho_{k, m, b}|$ for each mode.

%%%%%%%%%%%%%%%%%%%%%%%%%%%%%%%%%%%%%%%%%%%%%%%%%
\subsection{Computational Complexity}
\label{sssec:complexity}

According to \cite{de2015efficient}, the total number of FLOPS per FDN (without the direct path gain addition) is, 
\begin{equation}
\mathcal{O}_{\textrm{FDN}_k} = 2N'^2 + N'(P+1) + 2N'Q + 1, 
\end{equation}
where $2N'^2$ represents the FLOPS for matrix-vector multiplication in the feedback loop, $P$ is the number of operations required by the delay line gains/filters and $Q$ is the number of operations required by the source or receiver gains/filters. Now, for GFDNs with frequency-independent parameters, we simply have one multiplication per absorption gain and source-receiver gain.

Therefore, the total number of FLOPS per sample for processing $B$ parallel GFDNs with frequency-independent absorption gains and source-receiver gains is,
\begin{equation}
\begin{aligned}
\label{eq:FLOPS_single_gfdn_simplified}
\mathcal{O}_{\parallel \textrm{GFDN}} &= B \left(2N^2 + N(1+1) + 2N\right) + B \\
&= 2BN^2 + 4NB + B~. \\
\end{aligned}
\end{equation}
Here, the last $B$ operations denote the summation of $B$ DiffGFDN outputs. Note that additional FLOPS for the reconstructing filter bank processing need to be considered but they do not affect worst-case complexity. The filter bank is specified as $B$ FIR filters, and processes a buffer of input samples using fast FFT-based convolution. 

For minimal perceived colouration, thousands of modes are required in the FDN \cite{heldmann2021role}, with the minimally coloured FDNs having $10,000$ modes. For octave-band processing, i.e., $B=8$ with an average delay line length of $\tau_\text{ms} = 20$~ms at a sampling rate of $f_s = 44.1$~kHz, $12$ delay lines are adequate for generating over $10,000$~modes. The number of FLOPS required by the parallel subband GFDNs with these parameters is $2,696$.
% The FLOPS as a function of the number of delay lines for $B=8$ (modelling energy absorption and source-receiver EQs in octave bands) is shown in Fig.~\ref{fig:compare_FLOPS}. For upto $14$ delay lines, the parallel network outperforms the singular frequency-dependent GFDN.

In comparison, the CS model requires a total of $M_b \times B$ modes per group from (\ref{eq:rir_from_common_slopes_filter_form}), or a total of $G \times M_b \times B$ modes. Each mode can be reproduced by a biquad filter \cite{abel2014modal}, which requires $9$ FLOPS. The total number of FLOPS required by the static modal reverberators is $9 \times G \times M$, where $M = M_b \times B$. Similarly, each group requires an equaliser with $B$ cascaded biquads, so the total number of operations required by the EQs are, $9 \times G \times  B$. Therefore, the total number of FLOPS required by the proposed CS renderer in \cite{Gotz_Kerimovs_Schlecht_Pulkki} is $9 G \times(M + B)$. Karjalainen and Järveläinen \cite{karjalainen2001more} studied the number of modes, uniformly distributed over the Bark scale, required for perceptually diffuse reverberation, and used $5,000$ as a reference. For $G = 2, M=2,500$, the total number of modes is $5,000$ and the number of FLOPS is $45,144$ which exceeds the FLOPS required by the proposed GFDN architecture by an order of magnitude.

During rendering, as the source and listener move, the MLPs  output a new set of source-receiver gains. Let us assume that each DiffGFDN has a fully connected MLP with $N_{\textrm{layers}}$ hidden layers and $A$ neurons in each layer. Since the rate of movement is slower than the audio sampling rate, we will have to update the source-receiver filters/gains in a much slower thread. Nevertheless, with ReLU activation, the number of FLOPS per MLP for all the hidden layers is $N_{\textrm{layer}} (2A^2 + A)$. In the output layer, with $F$ learnable parameters, we have an additional set of operations which increases the total number of FLOPS to
\begin{equation}
\mathcal{O}_{\textrm{MLP}} = N_{\textrm{layer}}  (2A^2 + A) + (2AF + F).
\end{equation}

% The total number of operations related to the MLPs in the single DiffGFDN with source-receiver filters $F = 4GB$) is,
% \begin{equation}
% \mathcal{O}_{MLP_{GFDN}} = N_{layer}  (2A^2 + A) + (4 G B) (2A + 1),
% \end{equation}

\noindent The total number of operations related to all the MLPs in the parallel DiffGFDNs with source-receiver gains is
\begin{equation}
\mathcal{O}_{\textrm{MLP}_{\parallel \textrm{GFDN}}} = B \times \left(N_{\textrm{layer}} (2A^2 + A) + 2 G (2A + 1)\right).
\end{equation}

%%%%%%%%%%%%%%%%%%%%%%%%%%%%%%%%%%%%%%%%%%%%%%%%%%%%%%%%%%%%%%
\subsection{Memory Requirements}
\label{ssec:memory_requirements}

For the MLP parameterisation described above, the total number of learnable parameters that need to be stored per-MLP is
\begin{equation}
\label{num_params_MLP}
N_{\textrm{P}_\textrm{MLP}} = N_{\textrm{layer}}(A^2 + A) + 2G(A+1).
\end{equation} 
The other parameters in each DiffGFDN that need to be stored are $\sum_{i} m_i$ elements in the delay lines, $G$ common decay times, $N$ input-output gains, $N'^2$ feedback matrix elements, and $2G$ source-receiver gains. For an average delay line length of $\tau$ samples, $\sum_i m_i \approx \tau N$. Therefore, the memory requirement for a single frequency-independent DiffGFDN is
\begin{equation}
N_{\textrm{P}_\textrm{GFDN}} =N (\tau + 2) + 3G + N'^2. 
\end{equation}
The total memory requirement for $B$ parallel frequency-independent GFDNs with $B$ MLPs is
\begin{equation}
\begin{aligned}
N_{\textrm{P}_\textrm{total}} &= B \left(N_{\textrm{P}_\textrm{MLP}} + N_{\textrm{P}_\textrm{GFDN}} \right). \\
% &= B \left(N_{\textrm{layer}}(A^2 + A) + 2G(A+1)\right) + B \left(N(\tau + 2) + G + N'^2\right)
\end{aligned}
\end{equation}

% In Fig.~\ref{fig:compare_mems}, we plot the storage requirements in kilobytes (for $32$~bit floats) as a function of the number of delay lines for varying hyper-parameters in the MLP. The number of frequency bands, number of groups and average length of the delay lines are fixed at $B = 8, G = 2$ and $\tau_\text{ms}=20$~ms. 
The storage requirements increase linearly with the number of delay lines, and quadratically with the number of neurons in each layer. Even for deeper networks (e.g.: $N_\text{layer} = 10, A = 128$), a few megabytes are adequate for storing the parameters. In comparison, storing a dataset of RIRs over the entire spatial grid in a room would require gigabytes of memory.

\section{Evaluation}
\label{sec:eval}

We trained the proposed network on the publicly available three coupled room dataset \cite{Gotz_Kerimovs_Schlecht_Pulkki}, generated using Treble\footnote{\url{https://www.treble.tech/}}, which uses wave-based solvers up to $750$~Hz and ray tracing at higher frequencies. Each room has a uniform but distinct absorption coefficient ($\alpha_1 = 0.2$, $\alpha_2 = 0.01$, $\alpha_3 = 0.1$). The dataset includes $838$ receiver locations in the $(x, y)$ plane spaced $0.3$~m apart, all at a height of $1.5$~m, with a single source at $(2.0, 2.0, 1.5)$~m. RIRs are sampled at $32$~kHz and encoded in second-order ambisonics; we use only the omni-directional component.

Each GFDN in the parallel architecture has $N = 12$ delay lines grouped into $G = 3$ groups ($N' = 4$), 
% with co-prime lengths between $20–50$~ms for high modal density.
with co-prime lengths resulting in time delays between $20–50$~ms for high modal density.
We also experimented with $N=8$ delay lines per group, which yielded smoother reverberation at the cost of higher computational load. We report results with $N=4$ delay lines per group, as it is a more efficient structure that converged to similar loss values after optimisation.

With a fixed source, unit-magnitude source gains are used, and only the receiver gains are learned by the MLP. The source gain is kept fixed. The learnable matrices $\mathbf{W}_k$ (for the block-diagonal feedback matrix) are initialised with values from $\mathcal{U}[-\frac{1}{\sqrt{N}}, \frac{1}{\sqrt{N}}]$, and input/output gains from $\mathcal{U}[-\frac{1}{N}, \frac{1}{N}]$. The encoder uses spatial frequencies $f_{\textrm{min}} = 1$ m$^{-1}$, $f_{\textrm{max}} = 32$ m$^{-1}$, and $N_f = 20$ frequency components. Inputs are expanded to $B_b \times 6N_f$ for batch size $B_b = 32$. MLPs use ReLU activations, layer normalization \cite{lei2016layer}, and He initialization \cite{he2015delving}.

For subband training, RIRs were decomposed into octave bands using a $4096$-tap amplitude-preserving FIR filter bank \cite{antoni2010orthogonal}\footnote{Implemented in the Pyfar library: \url{https://pyfar.org}}. One DiffGFDN was trained per band, with scalar receiver gains learned by an MLP. Absorption gains were fixed based on the common decay times. We trained for $15$ epochs using the loss in (\ref{eq:total_loss}), with an EDC loss weighting of $\lambda_\text{EDC}=10$ and sparsity loss weighting of $\lambda_{\text{sparsity}}=2$. All other weights were set to $1$. MLP size varied by frequency band: $1$ hidden layer with $16$ neurons for $63$~Hz and $125$~Hz bands; $5$ hidden layers with $16$ neurons for $250$~Hz and $500$~Hz bands; and $3$ hidden layers with $128$ neurons for higher bands. These were found by searching over a grid of hyper-parameters. Input and output gains $\bm{b}$ and $\bm{c}$ were normalised at every step to ensure each group's impulse response has unit energy.\footnote{The code is available at \url{https://github.com/orchidas/DiffGFDN/} and sound examples are available at \url{https://ccrma.stanford.edu/~orchi/FDN/GFDN/DiffGFDN/}.}

\subsection{Ablation Studies}
\label{ssec:ablation}

\sisetup{
    reset-text-series = false, 
    text-series-to-math = true, 
    mode=text,
    tight-spacing=true,
    round-mode=places,
    round-precision=2,
    table-format=2.2,
    table-number-alignment=center
}
\begin{table*}
\caption{RMSE in octave-band EDC fits over receiver positions in the hold-out test set for different configurations tested in the ablation study. $80$\% of the remaining dataset was used for training in all cases.
\label{tab:ablation_results}}
\centering
\begin{tabular}{|l|*{8}{S}|}
\hline
\shortstack{Configuration} & 
\multicolumn{8}{|c|}{Mean subband DiffGFDN EDC error (dB) over receiver positions in test set} \\
% \shortstack{Subband GFDN \\ EDR error (dB)} \\
\hline
&  \text{63~Hz} & \text{125~Hz} & \text{250~Hz} & \text{500~Hz} & \text{1~kHz} & \text{2~kHz} & \text{4~kHz} & \text{8~kHz} \\
\hline
\textit{Full-band} & 8.9 & 4.40 & 4.13 & 5.13 & 4.81 & 1.88 & 2.09 & 3.11 \\
\hline
\textit{Colourless Prototype} & 8.44 & 3.97 & 4.09 &3.47 & 2.48 & 2.94 & 3.36 & 1.67 \\
\hline
\textit{Coupled} & 4.29 & 1.5 & 1.88 & 1.51 & 1.15 & 1.61 & 2.82 & 1.41 \\
\hline
\textit{Decoupled (Proposed)} & 4.13 & 1.54 & 1.49 & 1.50 & 1.62 & 1.87 & 2.09 & 1.46 \\
\hline
\end{tabular}
\end{table*}

To rigorously evaluate the architectural choices underlying the proposed model, we conducted a series of controlled ablation studies, each designed to isolate the effect of a specific modelling assumption. The number of delay lines was kept fixed across all configurations, as this parameter has minimal influence on the EDC mismatch and primarily affects perceptual attributes such as colouration and echo density. $10\%$ of the receiver positions in the dataset was used to create a hold-out test set. The remaining receiver positions in the dataset were split into $80\%-20\%$ for training and validation. The EDC errors are reported on the test set.

\begin{enumerate}
\item \textit{Full-band}:
The first experiment evaluates the viability of a unified broadband architecture. In this configuration, the MLP directly predicts the command gains and resonant frequencies of the biquad filters that implement the frequency-dependent receiver gains $g_{o,k}(z)$. Thus, only one broadband GFDN is employed, and its parameters, including the feedback matrix and input–output gains, are learned jointly with the MLP weights and biases. To give the model more expressive power, a single MLP with $10$ hidden layers and $64$ neurons per layer is used.
This ablation tests whether a single, more expressive model can replace the subband architecture without sacrificing accuracy.

\item \textit{Colourless Prototype}:
The second experiment examines whether subband-specific structural parameters are necessary. Here, we fixed the set of delay-line lengths across all subband GFDNs and applied colourless optimisation \cite{dal2023differentiable} to obtain a single shared set of feedback matrix, $\bm{A}$, and input-output gains, $\bm{b}, \bm{c}$. All subband GFDNs therefore possess identical colouration and echo density, and differ only through their receiver gains predicted by the MLP. An advantage of this configuration is that, after training, the MLP-predicted subband receiver gains can be consolidated into a single GEQ operating on octave-band cut-off frequencies, enabling inference with a single broadband GFDN. This study evaluates whether the benefits of subband decomposition arise from structural diversity across bands or merely from band-specific gain modulation.

\item \textit{Coupled}:
The third experiment investigates the role of inter-group decoupling in the feedback structure. In this variant, we relaxed the block-diagonal constraint on the feedback matrix by allowing non-zero off-diagonal entries, yielding a fully learnable dense feedback matrix. This modification introduces inter-group coupling in the feedback loop and breaks the correspondence with the CS formulation. 
This ablation assesses whether explicit inter-group coupling enhances modelling accuracy or instead degrades stability and interpretability relative to the proposed decoupled structure.
\end{enumerate}

Table~\ref{tab:ablation_results} summarises the subband EDC mismatch errors obtained on the test set for all configurations and compares them with the performance of the final architecture adopted in this work, referred to as \textit{Decoupled}, which clearly outperforms \textit{Colourless Prototype} and \textit{Full-band} in all octave bands. \textit{Decoupled} gives better or similar EDC fits compared to \textit{Coupled} in all bands except $1$~kHz. For the \textit{Coupled} architecture, the learnable coupling matrix becomes more diagonal after optimisation, i.e., starting from a dense matrix, the optimisation makes the final structure more closely resemble a diagonal matrix. This was verified by computing the energy ratio of the diagonal elements to its total energy pre- and post-optimisation. While allowing inter-group coupling can give marginal benefits, it breaks the correspondence with the CS model, and increases computational complexity significantly.

% \subsection{Parallel subband DiffGFDN training in octave bands}
% \label{ssec:treble_subband_training}

%%%%%%%%%%%%%%%%%%%%%%%%%%%%%%%%%%%%%%%%%%%%
% \begin{figure}
% \centering
% \subfloat[Amplitudes derived from DiffGFDN RIRs at $1$~kHz as a function of spatial location]{
%  \includegraphics[trim=100 0 0 0, clip,width=0.22\textwidth]{Figures/treble_subband_training/treble_data_grid_training_1000Hz_colorless_loss_diff_delays.yml_epoch=14_src=(2.00, 2.00, 1.50)_learnt_amplitudes_in_space.png}
% \label{fig:treble_data_grid_training_diff_gfdn_amps}
% } \quad
% \subfloat[Common slope amplitudes estiamted using least squares at $1$~kHz as a function of spatial location]{
%  \includegraphics[trim=100 0 0 0, clip,width=0.22\textwidth]{Figures/treble_subband_training/treble_data_grid_training_1000Hz_colorless_loss_diff_delays.yml_epoch=14_src=(2.00, 2.00, 1.50)_actual_amplitudes_in_space.png}
% \label{fig:treble_data_grid_training_common_slopes_amps}
% }
%  \caption{Amplitudes corresponding to the three slopes for DiffGFDN (left) and common slopes model (right) for a subband frequency of $1000$~Hz. The spatial variations of the amplitudes in the common slope model are much higher. The source position is marked with a red cross.}
% \end{figure}

%%%%%%%%%%%%%%%%%%%%%%%%%%%%%%%%%%%%%%%%%%%%%%%%%%%%%%%%%%%
\subsection{Evaluation of Decoupled Subband DiffGFDNs}
\subsubsection{Spectral Colouration and Echo Density}
\label{sssec:subband_training_colouration_loss}

\begin{figure*}
\centering
\subfloat[Pre-optimisation magnitude spectrum of a group in the GFDN]{
 \includegraphics[width=0.45\textwidth]{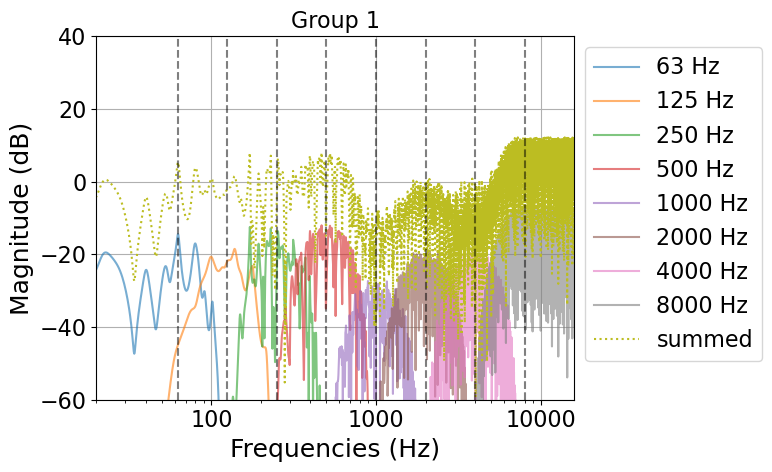}
\label{fig:init_gfdn_mag_spec}
} \qquad
\subfloat[Post-optimisation magnitude spectrum of a group in the GFDN]{
 \includegraphics[width=0.45\textwidth]{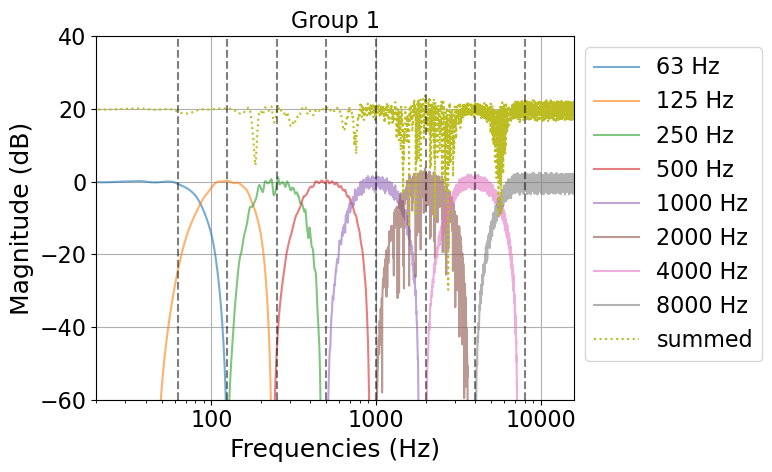}
\label{fig:opt_gfdn_mag_spec}
}
\caption{ The left plot shows the magnitude response of one group in the GFDN before training, and the right plot shows the magnitude response of the same group post-training. Before training, the GFDN has random input-output gains and a random unitary feedback matrix; the training tunes these parameters for minimal colouration. A flat magnitude response is desirable. The different colours show the magnitude response in different subbands, the magnitude response summed over all subbands is offset by $20$~dB and shown in yellowish green dotted lines.}
\label{fig:mag_spectrum}
\end{figure*}

\begin{figure*}
\centering
\subfloat[NED for $f_c = 63$~Hz]{
 \includegraphics[width=0.21\textwidth, height=2.5cm]{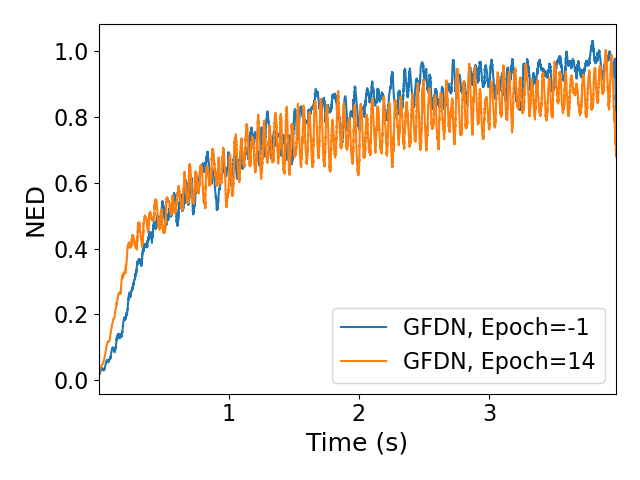}
\label{fig:treble_data_grid_training_ned_125Hz}
} \hspace{0.02\textwidth}
\subfloat[NED for $f_c = 125$~Hz]{
 \includegraphics[width=0.21\textwidth, height=2.5cm]{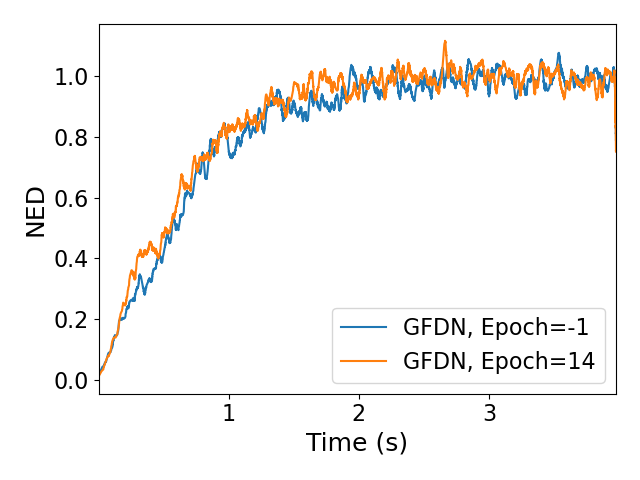}
\label{fig:treble_data_grid_training_ned_125Hz}
} \hspace{0.02\textwidth}
\subfloat[NED for $f_c = 250$~Hz]{
 \includegraphics[width=0.21\textwidth, height=2.5cm]{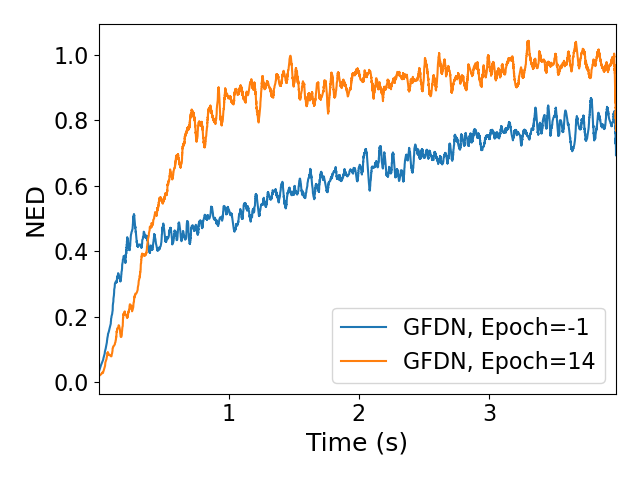}
\label{fig:treble_data_grid_training_ned_250Hz}
} \hspace{0.02\textwidth}
\subfloat[NED for $f_c = 500$~Hz]{
 \includegraphics[width=0.21\textwidth, height=2.5cm]{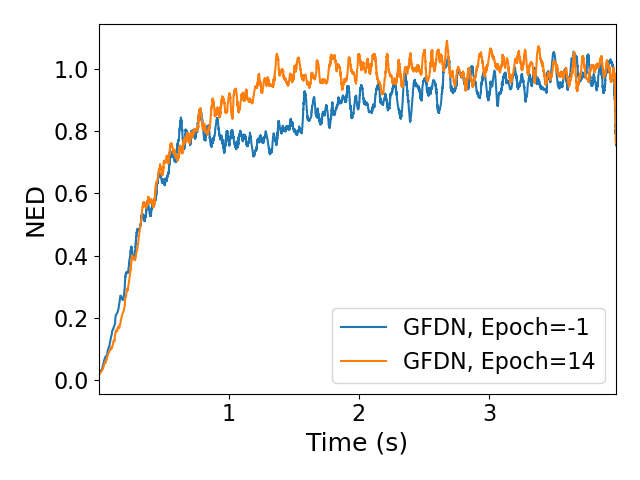}
\label{fig:treble_data_grid_training_ned_500Hz}
} \\
\subfloat[NED for $f_c = 1000$~Hz]{
 \includegraphics[width=0.21\textwidth, height=2.5cm]{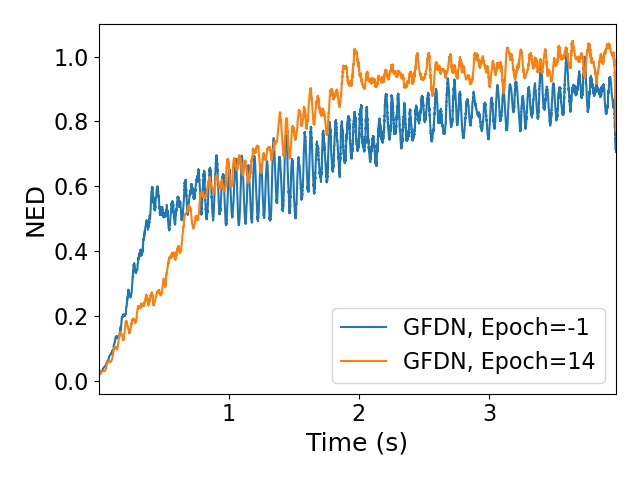}
\label{fig:treble_data_grid_training_ned_1000Hz}
}
\hspace{0.02\textwidth}
\subfloat[NED for $f_c = 2000$~Hz]{
 \includegraphics[width=0.21\textwidth, height=2.5cm]{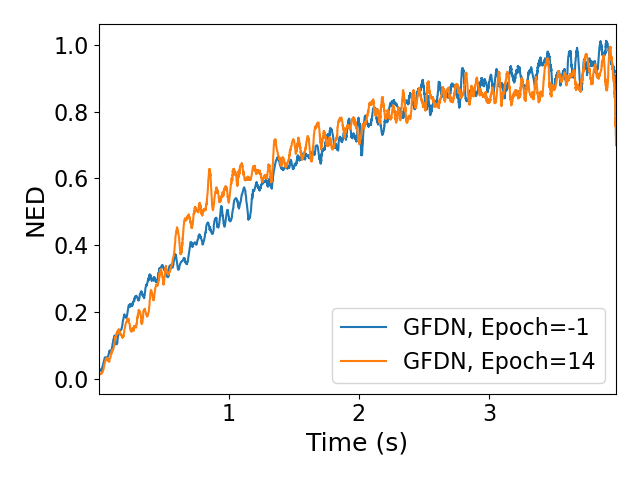}
\label{fig:treble_data_grid_training_ned_2000Hz}
} \hspace{0.02\textwidth}
\subfloat[NED for $f_c = 4000$~Hz]{
 \includegraphics[width=0.21\textwidth, height=2.5cm]{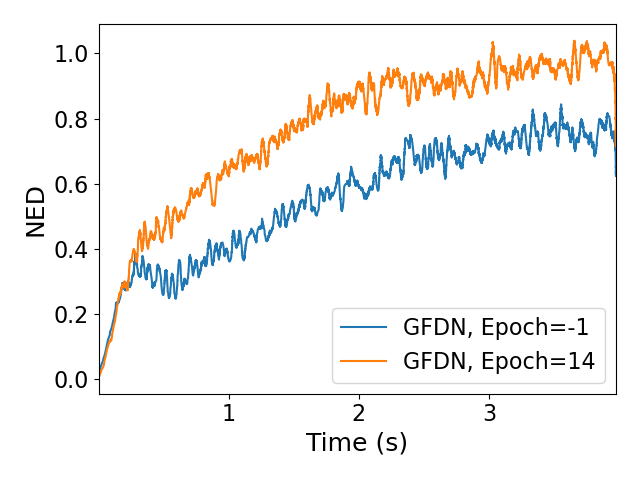}
\label{fig:treble_data_grid_training_ned_4000Hz}
} \hspace{0.02\textwidth}
\subfloat[NED for $f_c = 8000$~Hz]{
\includegraphics[width=0.21\textwidth, height=2.5cm]{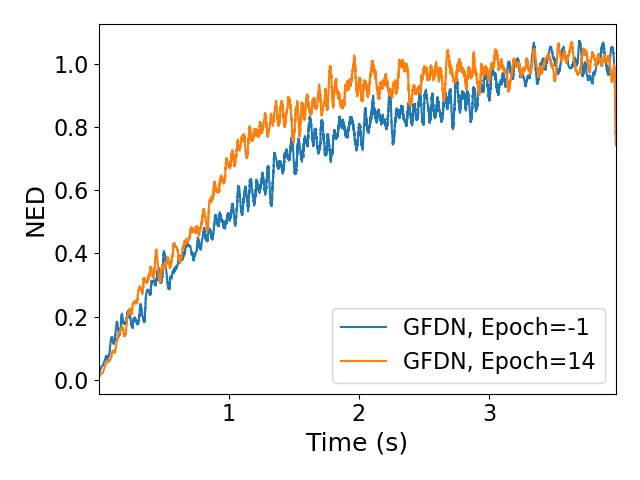}
\label{fig:treble_data_grid_training_ned_8000Hz}
}
 \caption{Comparison of NEDs at position (9.3, 6.6, 1.5)~m before (blue) and after training (orange) for different subband GFDNs. A faster rise in NED is desired for smoother reverberation.}
 \label{fig:treble_data_grid_training_ned_var_freqs}
\end{figure*}

For the $b^\text{th}$ frequency band, the lossless transfer function of the $k^\text{th}$ group is
\begin{equation}
\label{eq:sub_fdn_mag_res}
\hat{H}_{k, b}(z) = \bm{c}_{k, b}^T \left(\bm{D}_{\bm{m}_k}^{-1}(z) - \mathbf{M}_{k, b} \right)^{-1} \bm{b}_{k, b}.
\end{equation}
The frequency response of each group in the $b^\text{th}$ subband GFDN is then filtered with the $b^\text{th}$ subband filter's frequency response, $G_b(e^{j\omega})$, and summed to get the overall magnitude response of each group,
\begin{equation}
\label{eq:summed_mag_response}
    |\hat{H}_{k}(e^{j\omega})|^2 = \left|\sum_{b=1}^B \hat{H}_{k, b}(e^{j\omega}) G_b(e^{j\omega}) \right|^2.
\end{equation}
The magnitude responses, $|\hat{H}_{k}(e^{j\omega})|^2$, of a single group are offset by $20$~dB and shown in Fig.~\ref{fig:mag_spectrum} in yellowish-green dotted lines. Fig.~\ref{fig:init_gfdn_mag_spec} shows the magnitude response before training the parameters of the DiffGFDN, and Fig.~\ref{fig:opt_gfdn_mag_spec} shows the magnitude response post-training. Other colours on the plots show the magnitude response of each subband GFDN after filtering with the respective subband filter, i.e., $|\hat{H}_{k, b}(e^{j\omega}) G_b(e^{j\omega})|^2$. 

The magnitude response, $|\hat{H}_{k}(e^{j\omega})|^2$, should be as flat as possible if the spectral loss in (\ref{eq:spectral_loss}) has been minimised. We see that the group with learned parameters has a much flatter spectrum than the group with randomly initialised parameters. However, some notches are still present at the crossover frequencies after training.

To observe the sparsity losses pre- and post-training, we plotted the normalised echo density (NED) profile \cite{abel2006simple} for the receiver at $(9.3,6.6,1.5)$~m in Fig.~\ref{fig:treble_data_grid_training_ned_var_freqs}. The NED is a measure of the density of echo build-up across time, and should reach a value of $1$ for Gaussian noise. For all frequency bands shown in Fig.~\ref{fig:treble_data_grid_training_ned_var_freqs}, except $f_c=125$~Hz and $2000$~Hz, the rise in echo density is much faster post-training. This is desirable for smooth reverberation \cite{schlecht2016feedback}. These plots of Figs.~\ref{fig:mag_spectrum} and \ref{fig:treble_data_grid_training_ned_var_freqs} show that training is able to reduce both the spectral colouration and sparsity losses successfully.

%%%%%%%%%%%%%%%%%%%%%%%%%%%%%%%%%%%%%%%%%%%%%%%%

\subsubsection{Comparison with Common Slopes Model}
\label{ssec:comp_all_models}

%%%%%%%%%%%%%%%%%%%%%%%%%%%%%%%%%%%%%%%%%%%%%%%%%%%%%%%
\begin{figure*}[!t]
\centering
\subfloat[$f_c=63$ Hz, RMSE = $3.7$~dB]{
 \includegraphics[trim=80 0 0 0, clip,height=0.15\textwidth]{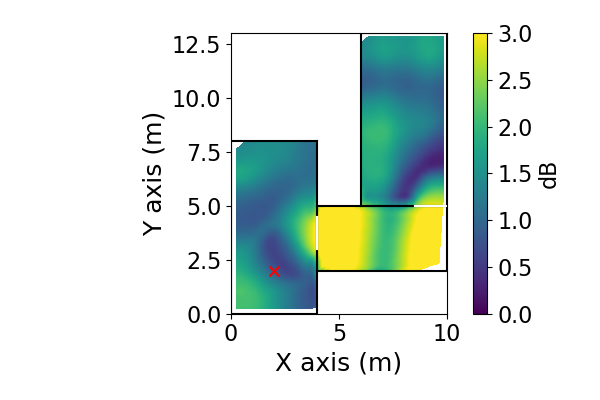}
\label{fig:treble_data_grid_training_common_slopes_edc_63}
} \hspace{0.01\textwidth}
\subfloat[$f_c=125$ Hz, RMSE = $0.7$~dB]{
 \includegraphics[trim=80 0 0 0, clip,height=0.15\textwidth]{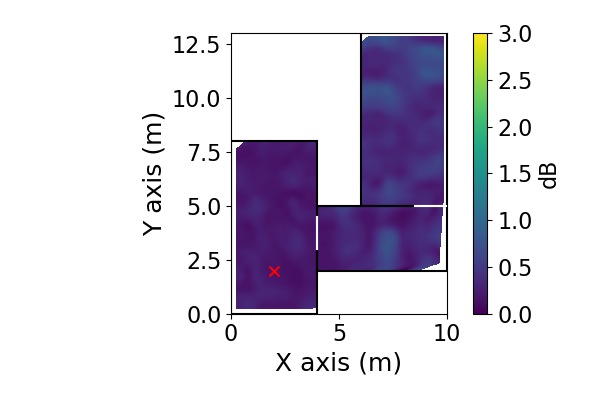}
\label{fig:treble_data_grid_training_common_slopes_edc_125}
} \hspace{0.01\textwidth}
\subfloat[$f_c=250$ Hz, RMSE = $0.4$~dB]{
 \includegraphics[trim=80 0 0 0, clip,height=0.15\textwidth]{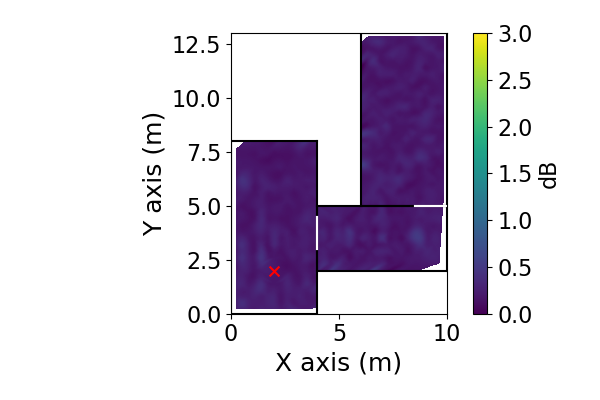}
\label{fig:treble_data_grid_training_common_slopes_edc_250}
} \hspace{0.01\textwidth}
\subfloat[$f_c=500$ Hz, RMSE = $0.4$~dB]{
 \includegraphics[trim=80 0 0 0, clip,height=0.15\textwidth]{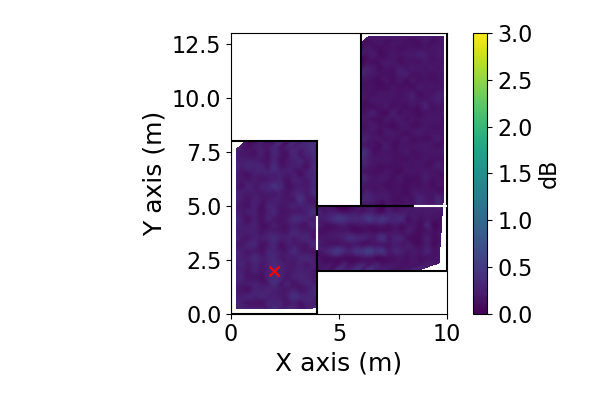}
\label{fig:treble_data_grid_training_common_slopes_edc_500}
} \\
\subfloat[$f_c=1$ kHz, RMSE = $0.3$~dB]{
 \includegraphics[trim=80 0 0 0, clip,height=0.15\textwidth]{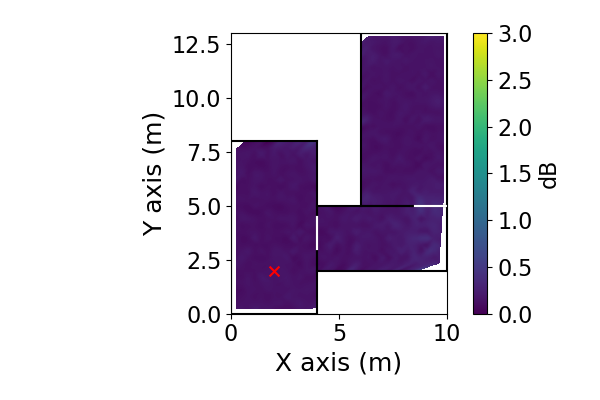}
\label{fig:treble_data_grid_training_common_slopes_edc_1k}
} \hspace{0.01\textwidth}
\subfloat[$f_c=2$ kHz, RMSE  = $0.3$~dB]{
 \includegraphics[trim=80 0 0 0, clip,height=0.15\textwidth]{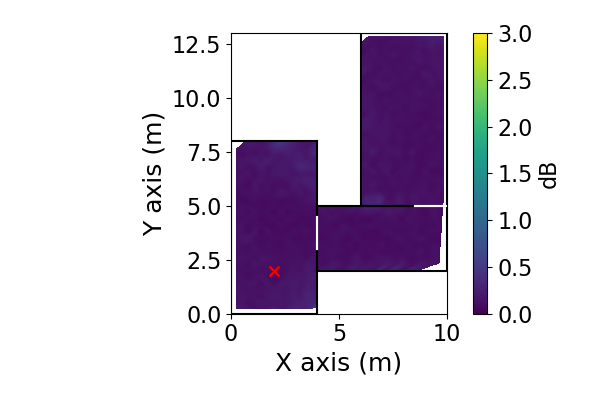}
\label{fig:treble_data_grid_training_common_slopes_edc_2k}
} \hspace{0.01\textwidth}
\subfloat[$f_c=4$ kHz, RMSE = $0.6$~dB]{
 \includegraphics[trim=80 0 0 0, clip,height=0.15\textwidth]{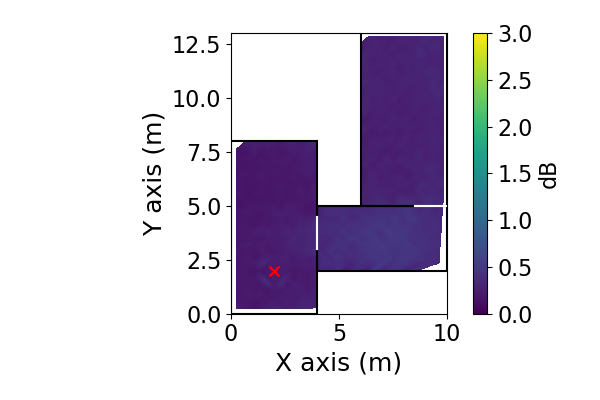}
\label{fig:treble_data_grid_training_common_slopes_edc_4k}
} \hspace{0.01\textwidth}
\subfloat[$f_c=8$ kHz, RMSE = $0.9$~dB]{
 \includegraphics[trim=80 0 0 0, clip,height=0.15\textwidth]{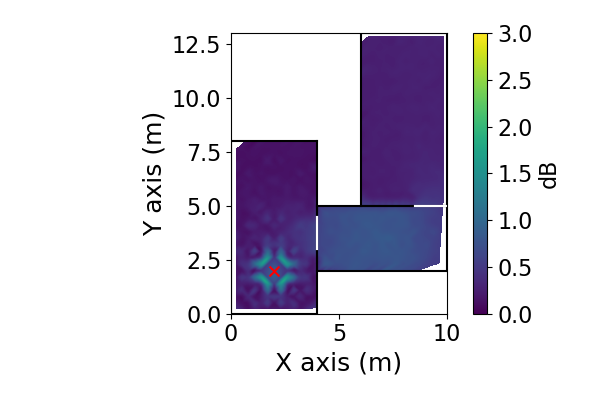}
\label{fig:treble_data_grid_training_common_slopes_edc_8k}
}
 \caption{Mean EDC fit error between the original and CS synthesised RIRs at all receiver positions in the Treble dataset. An error of $0$~dB indicates a perfect match. The source position is marked with a red cross. The CS model required an RIR measurement at all positions.}
\label{fig:subband_edc_error_common_slopes}
\end{figure*}
%%%%%%%%%%%%%%%%%%%%%%%%%%%%%%%%%%%%%%%%%%%%%%%%%%%%%%%%%%%%%%%%%
\begin{figure*}[!htbp]
\centering
\subfloat[$f_c=63$ Hz, RMSE = $5.0$~dB]{
 \includegraphics[trim=80 0 0 0, clip, height=0.15\textwidth]{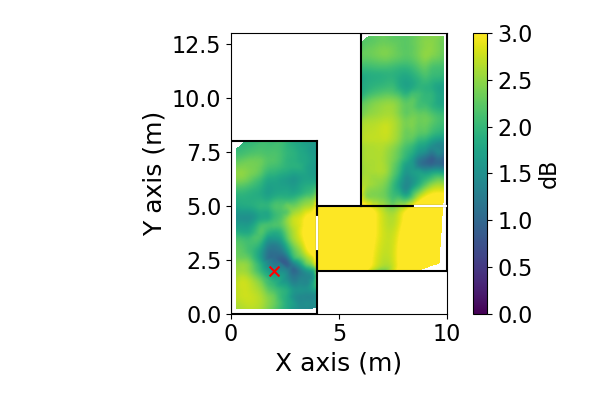}
\label{fig:treble_data_grid_training_diff_gfdn_edc_63}
} \hspace{0.01\textwidth}
\subfloat[$f_c=125$ Hz, RMSE = $1.4$~dB]{
 \includegraphics[trim=80 0 0 0, clip,height=0.15\textwidth]{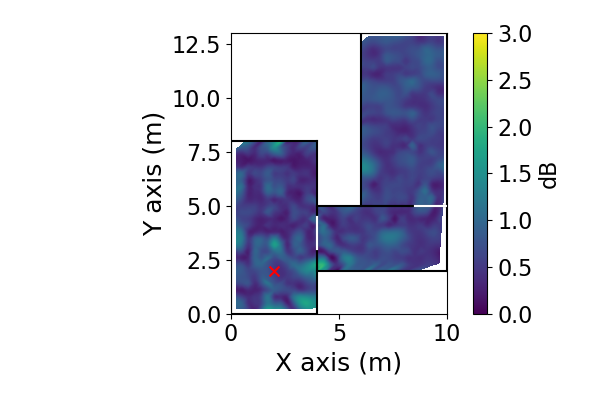}
\label{fig:treble_data_grid_training_diff_gfdn_edc_125}
} \hspace{0.01\textwidth}
\subfloat[$f_c=250$ Hz, RMSE = $1.2$~dB]{
 \includegraphics[trim=80 0 0 0, clip,height=0.15\textwidth]{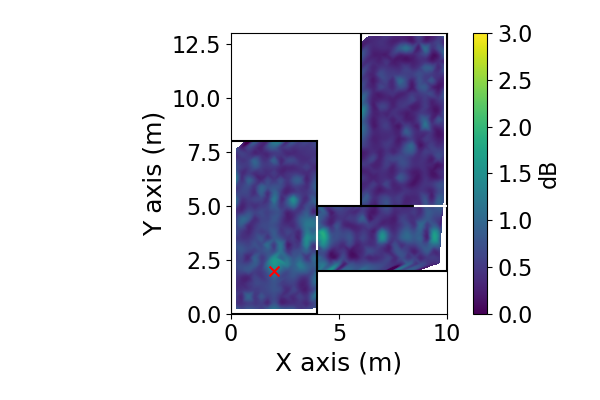}
\label{fig:treble_data_grid_training_diff_gfdn_edc_250}
} \hspace{0.01\textwidth} 
\subfloat[$f_c=500$ Hz, RMSE = $1.5$~dB]{
 \includegraphics[trim=80 0 0 0, clip,height=0.15\textwidth]{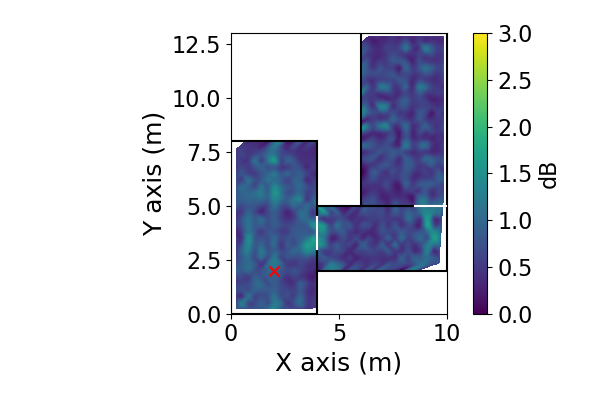}
\label{fig:treble_data_grid_training_diff_gfdn_edc_500}
} \\
\subfloat[$f_c=1$ kHz, RMSE = $0.8$~dB]{
 \includegraphics[trim=80 0 0 0, clip,height=0.15\textwidth]{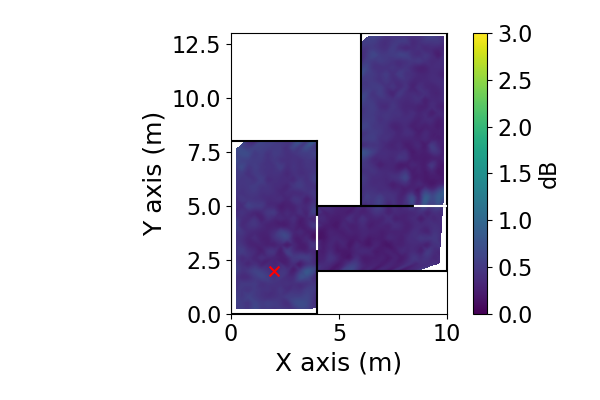}
\label{fig:treble_data_grid_training_diff_gfdn_edc_1k}
} \hspace{0.01\textwidth}
\subfloat[$f_c=2$ kHz, RMSE  = $1.2$~dB]{
 \includegraphics[trim=80 0 0 0, clip,height=0.15\textwidth]{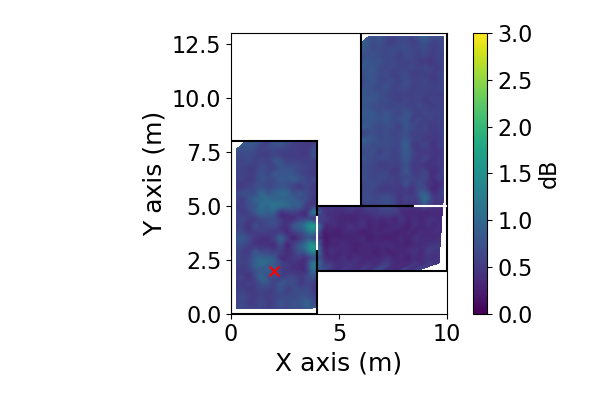}
\label{fig:treble_data_grid_training_diff_gfdn_edc_2k}
} \hspace{0.01\textwidth}
\subfloat[$f_c=4$ kHz, RMSE = $1.7$~dB]{
 \includegraphics[trim=80 0 0 0, clip,height=0.15\textwidth]{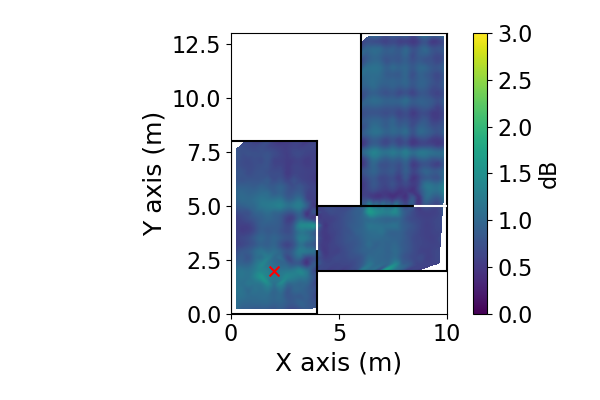}
\label{fig:treble_data_grid_training_diff_gfdn_edc_4k}
} \hspace{0.01\textwidth}
\subfloat[$f_c=8$ kHz, RMSE = $1.4$~dB]{
 \includegraphics[trim=80 0 0 0, clip,height=0.15\textwidth]{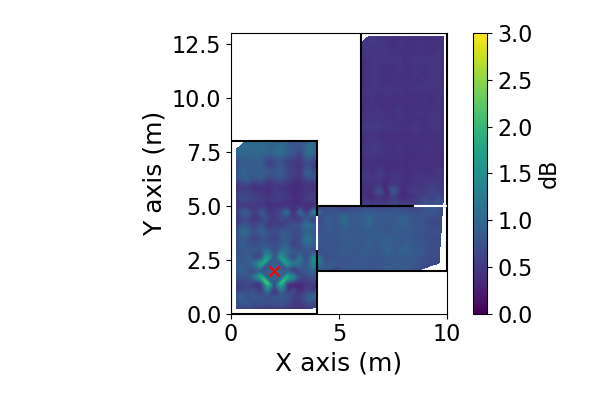}
\label{fig:treble_data_grid_training_diff_gfdn_edc_8k}
}
\caption{Mean EDC fit error between the original and proposed parallel subband DiffGFDN synthesised RIRs, which required training using a subset of positions only, cf.~Fig.~\ref{fig:subband_edc_error_common_slopes}.}
\label{fig:subband_edc_error_diff_gfdn_subband}
\end{figure*}

%%%%%%%%%%%%%%%%%%%%%%%%%%%%%%%%%%%%%%%%%%%%%%%%%%%%%
\begin{figure}
\centering
\subfloat[Broadband EDC fits for receiver at ($6.4, 3.8, 1.5$)~m]{
 \includegraphics[trim=0 0 0 30, clip, width=0.45\textwidth, height=5.5cm]{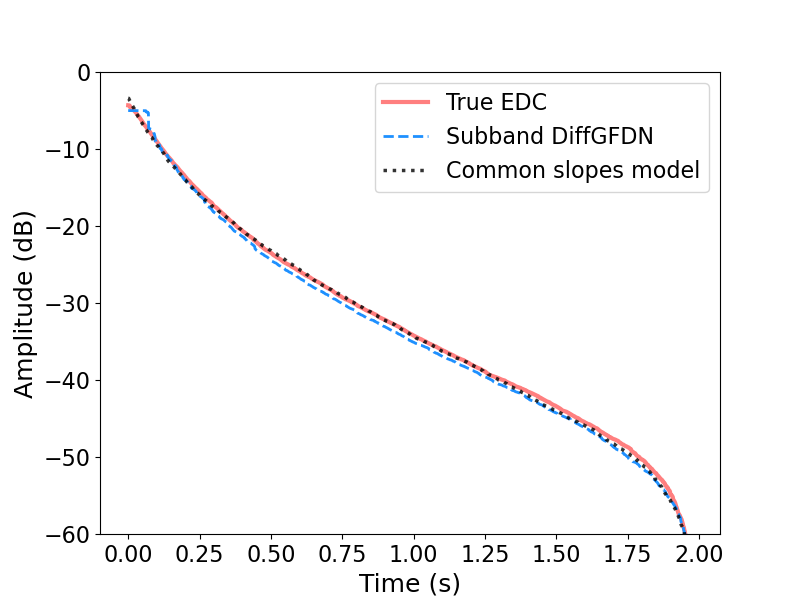}
\label{fig:broadband_edc_pos1}
}\\
\subfloat[Broadband EDC fits for receiver at ($9.3, 6.6, 1.5$)~m]{
 \includegraphics[trim=0 0 0 30, clip, width=0.45\textwidth, height=5.5cm]{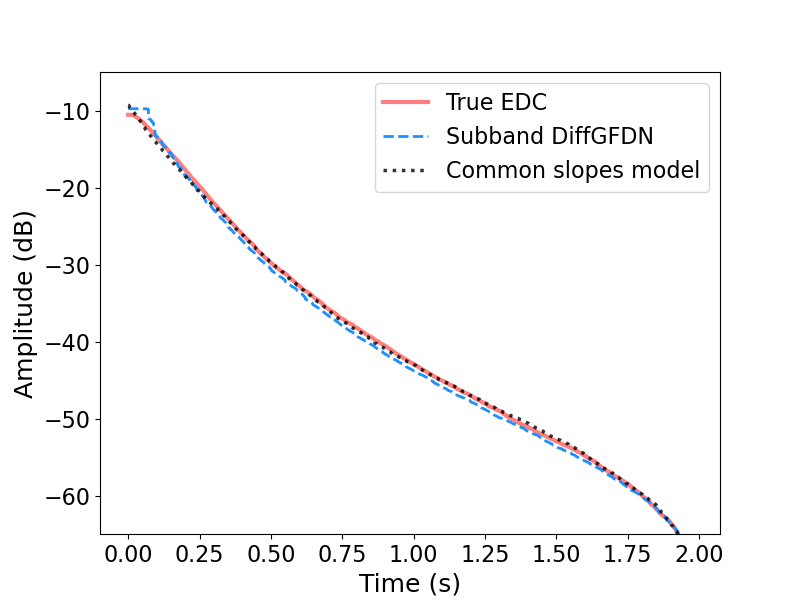}
\label{fig:broadband_edc_pos2}} 
 \caption{{Comparison of broadband EDC fits of the proposed subband GFDN architecture with the CS model and ground truth, for two receiver positions in the dataset. The multi-slope decay is evident in both plots.}}
 \label{fig:compare_edc_all_models}
\end{figure}
%%%%%%%%%%%%%%%%%%%%%%%%%%%%%%%%%%%%%%%%%%%%%%%%%%%%%%%%%%%%%%%%%%%%%%%%%%%%%%%%%%%%%%%%%%%%%%%%%%%%%%%%

\begin{figure}
\centering
\subfloat[Subband DiffGFDN, RMSE = $3.2$~dB]{
 \includegraphics[trim=50 10 20 10, clip, width=0.45\columnwidth]{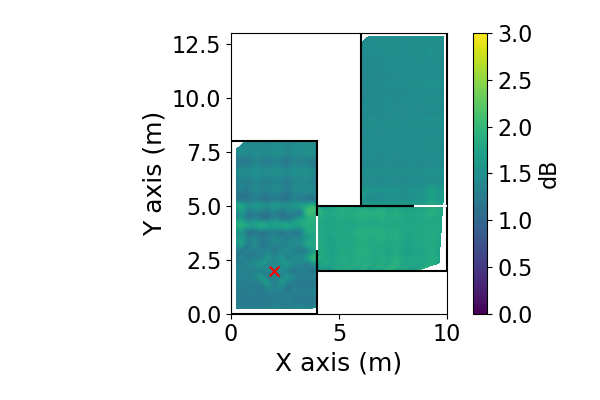}
\label{fig:treble_data_grid_training_single_edr_comp_subband_gfdn}
}\quad
\subfloat[CS model, RMSE = $4.8$~dB]{
 \includegraphics[trim=50 10 20 10, clip, width=0.45\columnwidth]{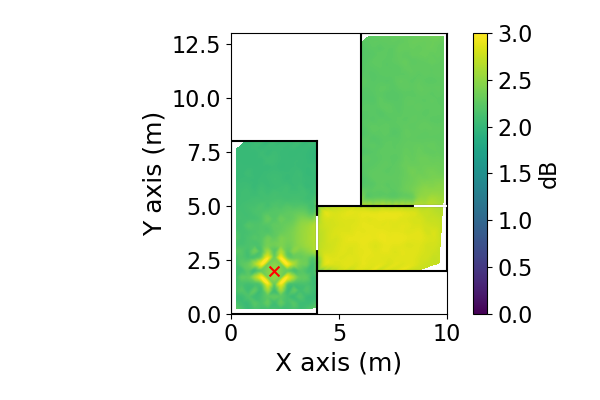}
\label{fig:treble_data_grid_training_overall_edr_comp_common_slopes}} 
 \caption{Comparison of mean absolute EDR fit errors of the proposed method and the CS model when the training set contains $80\%$ of the receivers.}
 \label{fig:treble_data_grid_training_edr_comp_all_models}
\end{figure}

%%%%%%%%%%%%%%%%%%%%%%%%%%%%%%%%%%%%%%%%%%%%%%%%%%%%%%%%%%%
\begin{itemize}

\item \textit{Octave Band EDC Losses}:
\label{sssec:octave_band_edc_loss}
To analyse the results on the entire grid of positions, we plotted the error in EDC matching between the reference RIRs and the DiffGFDN's RIRs for each position and each frequency-band. A total of $80\%$ of the $838$ receiver positions were randomly chosen to be in the training set. The plots show the EDC error for all available receiver positions in the dataset. This band-wise EDC error is calculated as
\begin{equation}
\label{eq:edc_error}
\epsilon_{\text{EDC}_b}(\mathbf{x}) = \frac{1}{Q} \sum_{n=1}^{Q} \left|d_{b_\text{dB}}(\mathbf{x},n) - \hat{d}_{b_\text{dB}}(\mathbf{x}, n) \right|,
\end{equation}
where $d_{b_\text{dB}}$ is the EDC of the true RIR in the $b^\text{th}$ frequency-band, $\hat{d}_{b_\text{dB}}$ is the EDC of the synthesized RIR in the $b^\text{th}$ frequency-band in decibels and $Q$ is the total number of time samples. 

Figs.~\ref{fig:subband_edc_error_common_slopes} and \ref{fig:subband_edc_error_diff_gfdn_subband} show the mean absolute EDC error, $\epsilon_{\text{EDC}_b}(\mathbf{x})$, as a function of receiver position according to (\ref{eq:edc_error}) for the CS model and the proposed subband DiffGFDN architecture in octave bands. The common decay times in the CS model are the same as those used to derive the absorption gains in the DiffGFDNs. The position-dependent CS amplitudes are derived using non-linear least squares by minimising the difference between the EDCs obtained from the reference RIRs and CS predicted EDCs. Note that the CS model requires measured RIRs at all positions in the dataset, whereas the DiffGFDNs were trained on RIRs from a subset of positions (training set), and interpolated the RIRs at other positions (test set). Although the subband DiffGFDN yields consistently low EDC errors below $1.8$~dB in Fig.~\ref{fig:subband_edc_error_diff_gfdn_subband}, the CS model gives a better root-mean-squared-error (RMSE) in all frequency-bands, in Fig.~\ref{fig:subband_edc_error_common_slopes}.
Both methods perform poorly in the $63$~Hz band, where dominant room modes need to be modelled individually.

\item \textit{EDR Losses}:
\label{sssec:wide_band_losses}
The final RIRs from the subband DiffGFDNs are synthesised by passing an impulse through the trained subband DiffGFDNs in parallel, and filtering the output of each DiffGFDN with the appropriate subband filter, before summing all the frequency bands. The CS model's RIRs are synthesised by shaping white noise in octave bands. The slope of the RIRs in each octave band is determined by the common decay times, and the initial level of the EDC in each octave-band for each RIR is determined by the position-dependent amplitudes.

In Fig.~\ref{fig:compare_edc_all_models}, we show the EDC plots for the broadband RIR reconstructed from the subband GFDN outputs and the CS parameters against the ground truth data for two receiver positions in the second and third room respectively. The plots clearly demonstrate multi-slope decay, which is captured accurately by the proposed subband GFDN architecture.

The mean absolute EDR error as a function of spatial location is shown in Fig.~\ref{fig:treble_data_grid_training_edr_comp_all_models}. This error is calculated as
\begin{equation}
 \epsilon_{\text{EDR}}(\mathbf{x}) = \frac{1}{K} \sum_k \frac{1}{J} \sum_j |\text{EDR}(\mathbf{x}, k, j) - \widehat{\text{EDR}}(\mathbf{x}, k, j)|.
\end{equation}
 The subband DiffGFDN architecture, which has an RMSE of $3.2$~dB, outperforms the CS model, which has an RMSE of $4.8$~dB. The CS model shows better performance in terms of the EDC error, but has a higher overall EDR error. While this may appear confounding, the CS model gives exact fits to the EDCs at the octave frequencies, but its performance degrades in the frequencies between octave bands. In the subband GFDN, we explicitly minimise the EDR deviation from the reference, which leads to better performance.
\end{itemize}
\subsection{Spatial Sampling Requirements}
 \label{ssec:spatial_sampling}

\sisetup{
    reset-text-series = false, 
    text-series-to-math = true, 
    mode=text,
    tight-spacing=true,
    round-mode=places,
    round-precision=1,
    table-format=2.1,
    table-number-alignment=center
}
\begin{table*}
\caption{RMSE in octave-band EDC fits over receiver positions in the hold-out test set as a function of different number of receivers used in the training set. The lowest errors in each frequency band are written in bold text.
\label{tab:table1}}
\centering
\begin{tabular}{|l|*{8}{S}|}
\hline
\shortstack{Number of \\measurements \\used in training} & 
\multicolumn{8}{|c|}{Mean subband DiffGFDN EDC error (dB) over receiver positions in test set} \\
% \shortstack{Subband GFDN \\ EDR error (dB)} \\
\hline
&  \text{63~Hz} & \text{125~Hz} & \text{250~Hz} & \text{500~Hz} & \text{1~kHz} & \text{2~kHz} & \text{4~kHz} & \text{8~kHz} \\
\hline
10\% \ (76) & 5.9 & 3.7 & 3.4 &3.5 & 2.9 & 3.9 & 4.2 & 1.7 \\
\hline
20\% \ (151) & 4.2 & 3.5 & 2.3 & 4.3 & 2.6 & 4.4 & 3.7 & 1.85 \\
\hline
30\% \ (227) & 
4.2 & 2.7 & 2.5 & 4.1 & 1.6 & 4.2 & 3.8 & 1.4\\
\hline
40\%  \ (302) & 3.8 & 3.2 & 1.8 & 2.5 & 1.7 & 4.2 & 3.9 & 1.6 \\
\hline
50\% \ (378) & 4.8 & 2.2 & 1.8 & 2.2 & 1.8 & 3.5 & 3.7 & 1.6\\
\hline
60\% \ (453) & 4.0 & 2.1 & 1.5 & 1.7 & 1.7 & \textbf{1.2} & \textbf{1.7} & \textbf{1.3} \\
\hline
70\% \ (529) & 4.0 & 1.9 & 1.4 & 1.7 & 2.3 & 1.7 & 1.5 & 1.4\\
\hline
80\% \ (604) & 4.3 & \textbf{1.7} & \textbf{1.3} & 1.6 & \textbf{1.3} & 3.8 & 1.8 & 1.4 \\
\hline
90\% \ (680) & \textbf{3.9} & 2.0 & 1.6 & \textbf{1.5} & 1.8 & 1.4 & 1.8 & 1.4\\
\hline
\end{tabular}
\end{table*}

 The effect of spatial sampling on the EDC RMSE is shown in Table~\ref{tab:table1}. A fixed hold-out test set comprising $10\%$ of the data was reserved for evaluation, while the remaining $90\%$ of the receiver positions were randomly split into training sets of varying sizes. For example, a $40\%$ split indicates that $40\%$ of the remaining data was used for training and the rest for validation. Training was run for $20$ epochs on an NVIDIA V100 GPU. All reported errors are computed over the fixed $10\%$ test set. Errors typically decrease with more training data, as expected. 
 Low-frequency bands, particularly $63$~Hz, exhibit the highest errors and remain difficult to learn even with large amounts of data, whereas mid-frequency bands ($250$~Hz, $500$~Hz, $1$~kHz) show steady improvement as more training RIRs are provided. High-frequency bands converge earlier, although the $2$~kHz and $4$~kHz bands display non-monotonic behaviour and require at least $60\%$ of the training data before stabilising. When using $80\%$ of the data for training, the error for the $2$~kHz band fails to converge within $20$ epochs. Across most octave bands, a noticeable reduction in error variability occurs around the $60\%$ mark, after which gains become more incremental. 
 
 % Different numbers and random distributions of receivers positions are used for training.  For example, a train-test split of $40\%$ would mean $40\%$ of the receivers positions were in the training set, and $60\%$ were in the test set. The training is run for $20$ epochs on an NVIDIA V100 GPU. The errors are calculated for receiver positions in the test sets. In general, EDC errors decrease when more RIRs are available for training, although not significantly. 

 % TO-BE-CHANGED
%  The model is able to learn the amplitude mappings even when only $250$ receiver positions are used for training, with a mean EDC error of $1.6$~dB. There is a noticeable improvement in the mid- and high-frequency bands EDC fits when the number of receivers in the training set is increased from from $50\%$ to $60\%$. 
% For all frequency bands except $1$~kHz and $2$~kHz, increasing the number of receivers in the training set from $80\%$ to $90\%$ results in decreased performance, likely due to overfitting. 
\section{Discussion}
\label{sec:discuss}

Ablation studies showed that the decoupled subband DiffGFDN architecture provides the best trade-off between accuracy and model complexity. Analysis of the magnitude response and NED curves further confirmed that jointly learning the feedback matrix and the input–output gains reduces timbral colouration and increases echo density. Compared to the state-of-the-art CS model, the proposed architecture yields slightly higher EDC errors but achieves lower EDR errors, while also enabling interpolation to unseen receiver positions.

We examined the effect of spatial sampling density on octave-band EDC errors using a fixed hold-out test set. As expected, using more RIR samples for training reduced the error which stabilised to values less than $3$~dB in all except the $63$~Hz band when using $60\%$ or more of the data for training. The only anomaly is the error for the $2$~kHz band when trained using $80\%$ of the data. Determining the spatial sampling requirements for perceptually robust late-reverberation modelling—including the number and spatial arrangement of necessary RIRs—remains an open research question.

A natural next step is to evaluate DiffGFDN on real measured rooms. However, the scarcity of high-quality multichannel RIR datasets poses practical challenges. Real measurements often contain noise and artefacts that are absent in simulated datasets. To operate reliably on such data, DiffGFDN would need to explicitly model the noise floor and potentially incorporate denoising into the training process. Addressing these issues and collecting suitable measurement datasets will be essential for validating the method in real-world acoustic environments.

The subband DiffGFDN architecture is well-suited for rendering late reverberation in coupled spaces for AR applications when dense RIR measurements are available. Unlike storing full RIR datasets in the cloud, the learned DiffGFDN parameters can be stored directly on-device. Our recent extension of the method to directional late reverberation using spherical microphone array measurements enables binaural rendering \cite{das2026differentiable}. Rendering is orders of magnitude more efficient than time–space–varying convolution, making DiffGFDN a memory- and computation-efficient alternative to methods such as Neural Acoustical Fields \cite{luo2022learning} and Novel View Acoustic Synthesis \cite{chen2023novel}. Furthermore, the method is geometry-agnostic, relying only on RIR measurements and corresponding source–receiver positions. However, it does not yet generalise across different geometries. Few-shot generalisation has been explored for other neural RIR prediction methods \cite{su2022inras, ick2025direction, majumder2022few} and represents a promising direction for future work with DiffGFDN.

%%%%%%%%%%%%%%%%%%%%%%%%%%%%%%%%%%%%%%%%%%%%%%%%%%%%%%%%%
\section{Conclusion}
\label{sec:conclusion}

We proposed a Differentiable Grouped Feedback Delay Network (DiffGFDN) architecture for learning late reverberation in measured coupled spaces. The network parameters are optimised using perceptually motivated loss functions derived from a set of measured RIRs. An MLP learns the mapping between receiver (or source) positions and the corresponding DiffGFDN parameters. During inference, these parameters are updated in real-time as the listener moves, enabling spatial interpolation of late reverberation at unmeasured positions. This allows efficient, real-time rendering of late reverberation in any measured coupled space, making the method well-suited for AR and XR applications.

We showed that the RIRs generated by DiffGFDN are analogous to those produced by the CS model, which is a state-of-the-art method. Compared to the CS model, which relies on modal reverberators and equalisers, DiffGFDN is more computationally efficient and interpolates to unseen positions. We proposed a subband architecture with multiple parallel decoupled GFDNs, and evaluated it in an ablation study against other configurations, and against the CS model on a simulated three-room dataset. We also studied the effect of varying the amount of training data on the octave-band EDC errors in a fixed hold-out test set.

% Although the network produced slightly higher EDC errors than the CS model, it achieved lower EDR errors. Lower EDC errors were achieved when more RIRs were used for training.
Future work will explore generalisation to unseen geometries, using decimating filterbanks, spatial sampling requirements near room transitions and formal perceptual evaluations in real measured environments.
\section{Acknowledgments}
The authors wish to thank Georg Götz for the helpful and insightful discussions around the common slope model.

The Aalto University School of Electrical Engineering funded the work of the second author and supported it with the HUCE infrastructure. The computational resources were provided by the Aalto Science-IT project.
%%%%%%%%%%%%%%%%%%%%%%%%%%%%%%%%%%%%%
\appendices
\section{Pole Magnitudes in Each Frequency Band}
\label{app:pole_magnitudes}

The magnitude range of the poles in the $b^\text{th}$ frequency-band for the $k^\text{th}$ FDN is given by 
\cite{schlecht2019modal},
\small
\begin{equation}
\label{eq:mode_mag_range}
\min\left(|\bm{\gamma}_k(e^{j\angle\lambda_{m, k, b}})|^{\frac{1}{\bm{m}_k}} \right) \leq |\lambda_{m, k, b}| \leq \max  \left(|\bm{\gamma}_k(e^{j\angle\lambda_{m, k, b}})|^{\frac{1}{\bm{m}_k}} \right).
\end{equation}
\normalsize
Note that $\bm{\gamma}_k(e^{j\angle\lambda_{m, k, b}}) \in \mathbb{C}^{\frac{N}{N_{G}} \times 1}$, and $\bm{m}_k \in \mathbb{Z}^{+\frac{N}{N_{G}} \times 1}$. Since these are vectors, all the operations here are done element-wise.

Let us assume we have a flat absorption filter response with a small  ripple $\delta >0$ for $\bm{\gamma}_k(z)$  in the $b^\text{th}$ frequency band, i.e., 
\footnotesize
\begin{equation*}
\begin{aligned}
&\bm{\gamma}_{k, b} -  \delta \leq |\bm{\gamma}_k(e^{j\angle\lambda_{{m, k, b}}})| \leq \bm{\gamma}_{k, b} +  \delta, \ \forall \ \frac{\angle \lambda_{m, k, b} f_s} {2\pi}  \in [\frac{f_b}{\sqrt{2}}, \sqrt{2}f_b],
\end{aligned}
\end{equation*}
\normalsize
where  $\gamma_{k, b}$ is the absorption gain for the $b^\text{th}$ frequency band, from (\ref{eq:decay_time_to_gains}). Using this inequality range,  we can write
\footnotesize
\begin{equation}
\bm{\gamma}_{k, b}^{\frac{1}{\bm{m}_k}} \left(1-\frac{\delta}{\bm{\gamma}_{k, b} }\right)^{\frac{1}{\bm{m}_k}}
    \leq  |\bm{\gamma}_k(e^{j\angle\lambda_{m, k, b}})^{\frac{1}{\bm{m}_k}}| \leq  
   \bm{\gamma}_{k, b} ^{\frac{1}{\bm{m}_k}}\left(1+\frac{\delta}{\bm{\gamma}_{k, b} }\right)^{\frac{1}{\bm{m}_k}} 
\end{equation}
\normalsize It follows that
\scriptsize
\begin{equation*}
\label{eq:mode_mag_range_simple}
 \min \left[  \bm{\gamma}_{k, b}^{\frac{1}{\bm{m}_k}} \left(1-\frac{\delta}{\bm{\gamma}_{k, b} }\right)^{\frac{1}{\bm{m}_k}} \right] \leq |\lambda_{m, k, b}| \leq 
 \max\left[
\bm{\gamma}_{k, b}^{\frac{1}{\bm{m}_k}} \left(1+\frac{\delta}{\bm{\gamma}_{k, b} }\right)^{\frac{1}{\bm{m}_k}}
 \right].
\end{equation*}
\normalsize From (\ref{eq:decay_time_to_gains}) we have, $\sqrt[\bm{m}_k]{\bm{\gamma}_{k, b}} = \exp \left( \frac{-6.91}{f_sT_{60_{k, b}}}\right)$. For sufficiently small $\delta / \bm{\gamma}_{k, b}$, (typically $<0.1$), and sufficiently large values of $\bm{m}_k$\footnote{For the constrained magnitude least squares method proposed in \cite{schlecht2017accurate}, the ripples, $\delta$, are typically in the range of $0-1$~dB, and the delay line lengths, $\bm{m}$ are in the range of $10-20$~ms ($400-1000$ samples at $f_s=44.1$~kHz) for adequate echo density.}, we can use the approximation, $|\lambda_{m, k, b}| \approx \sqrt[\bm{m}_k]{\bm{\gamma}_{k, b}} = |\lambda_{k, b}|$. 

\section{Modal Expansion With and Without Absorption Filters}
\label{app:modal expansion}

% Using (\ref{eq:modal_subband}), the spectral loss function can be written in terms of the sum of $M_b$ modes in each subband,
% \begin{equation}
% \begin{aligned}
% \mathcal{L}_{\text{spectral}_k} &= \frac{1}{L} \sum_{\ell=0}^{L-1}\left( \sum_{b=1}^{B} \sum_{m=1}^{M_b}\left |\frac{\rho_{m, k, b}}{1 - \tilde{\lambda}_{m, k, b}e^{-j\omega_\ell}}\right| - 1\right)^2 \\
% \end{aligned}
% \end{equation}
% Because we have ignored the absorption filters in (\ref{eq:spectral_loss}), note that $\tilde{\lambda}_{m, k, b} \neq \lambda_{m, k, b}$. However, the residues in a frequency band remain unchanged if the decay is approximately homogeneous in that frequency band.  This is shown below.

The transfer function of a lossless FDN prototype with $N$ delay lines can be written as a sum of $M$ modes 
\begin{equation}
\begin{aligned}
H_{\text{lossless}}(z) &= \bm{c}^T \left(\bm{D_{m}}^{-1}(z) - \mathbf{M} \right)^{-1} \bm{b}  \\
&= \sum_{m=1}^{M = \sum_i m_i} \frac{\rho_{m}}{1 - \tilde{\lambda}_{m} z^{-1}},
\end{aligned}
\end{equation}
where $\bm{b, c} \in \mathbb{R}^{N\times 1}$ are the input-output gains. $\bm{D_m}(z) = \text{diag}(z^{-m_1}, \ldots, z^{-m_N})$ is the diagonal matrix of delays, and $\mathbf{M} \in \mathbb{R}^{N \times N}$ is a unilossless feedback matrix \cite{schlecht2016lossless}. There are a total of $M =\sum_{i=1}^N m_i$ modes, the poles are $\tilde{\lambda}_m$, and the residues are $\rho_m$.

Note that the magnitude of the poles in this lossless FDN prototype is always 1, i.e., $|\tilde{\lambda}_{m}| = 1$. This can be derived from Rouche's theorem which implies  that $\min \left(\sigma(\mathbf{M}) \right)^{1/\bm{m}} \leq |\tilde{\lambda}_{m}| \leq \max \left(\sigma(\mathbf{M}) \right)^{1/\bm{m}}$ \cite{schlecht2019modal}, where $\sigma(\mathbf{M})$ denotes the singular values of the matrix $\mathbf{M}$. Because the matrix is designed to be unitary, the singular values are always $\pm 1$ and therefore $|\tilde{\lambda}_{m}| = 1$.

Now, if we have homogenous scalar attenuation in the FDN, that is represented by a diagonal matrix of attenuations, $\bm{\Gamma} = \text{diag}(\gamma^{\bm{m}})$, then we have a lossy FDN whose transfer function is 
\begin{equation}
\begin{aligned}
H_{\text{lossy}}(z) &= \bm{c}^T \left(\bm{D_{m}}^{-1}(z) \bm{\Gamma}^{-1} - \mathbf{M} \right)^{-1} \bm{b}_k  \\
&= \bm{c}_k^T  \left(\bm{D_{m}}^{-1}(\gamma^{-1} z) - \mathbf{M}\right)^{-1} \bm{b}\\
&= \sum_{m=1}^M \frac{\rho_{m}}{1 - \tilde{\lambda}_{m} \gamma z^{-1}} \\
&= \sum_{m=1}^M \frac{\rho_{m}}{1 - \lambda_{m} z^{-1}}
\end{aligned}
\end{equation}
Therefore, we see that the new poles, $\lambda_{m} = \gamma \tilde{\lambda}_{m}$ are just a scaled version of the original poles. The residues remain unaffected. 
% \textcolor{red}{ZC: that holds if in the line with delay $m_i$ samples your absorption is $\gamma^{m_i}$. Is that the case?}\textcolor{blue}{OD: Yes, that is the case.}

Now, for any general frequency-dependent attenuation, $\gamma(z)$, we have, $\bm{\Gamma}(z) = \text{diag}(\gamma(z)^{\bm{m}})$, and the transfer function is 
\begin{equation}
\begin{aligned}
H_{\text{lossy}}(z) &= \bm{c}^T \left(\bm{D_{m}}^{-1}(z) \bm{\Gamma}^{-1}(z) - \mathbf{M} \right)^{-1} \bm{b}  \\
&= \bm{c}^T  \left(\bm{D_{m}}(\gamma(z)^{-1} z) - \mathbf{M}\right)^{-1} \bm{b}\\
&= \sum_{m=1}^M \frac{\rho_{m}}{1 - \tilde{\lambda}_{m} \gamma (z) z^{-1}} \\
&= \sum_{m=1}^M \frac{\rho'_{m}}{1 - \lambda'_{m} z^{-1}}.
\end{aligned}
\end{equation}
Now, both the poles and residues change. However, if the $T_{60}$ is approximately constant in a frequency band, $b$, we have  $\gamma(z) = \gamma_b \forall \ \omega \in [\omega_{l_b}, \omega_{u_b}] $. In that case, the residues will remain the same in each frequency band, and we can write
\begin{equation}
\begin{aligned}
H_{\text{lossy}}(z) &= \sum_{b=1}^B \sum_{m=1}^{M_b} \frac{\rho_{m,b}}{1 - \tilde{\lambda}_{m,b} \gamma_b z^{-1}} \\
&= \sum_{b=1}^B \sum_{m=1}^{M_b} \frac{\rho_{m,b}}{1 - \gamma_b\tilde{\lambda}_{m,b} z^{-1}},
\end{aligned}
\end{equation}
where $\rho_{m,b}$ and $\tilde{\lambda}_{m,b}$ are the residues and poles in the $b^\text{th}$ frequency band. A practical design choice is to estimate the $T_{60}$ in octave bands and fit IIR filters to that with minimal deviation in the magnitude response between the centre frequencies; therefore, this approximation holds.

% \textcolor{red}{ZC: the assumption here is that in the line with delay $m_i$ samples your absorption is $\gamma(z)^{m_i}$. Is that the case? Also, if you have $\gamma(z)$ in the denominator, that would create different modes with different residuals.} \textcolor{blue}{OD: That's a very good point. The mode residues will change. Suppose the T60 is roughly constant in a frequency band, then $\gamma(z)$ is also constant in that band and the residues in that band will remain unchanged. In fact, Gloria did some simulations and plotted the mode residues for frequency-dependent attenuation, and saw that they were practically unchanged.}

\bibliographystyle{IEEEtran}
\bibliography{refs}

@inproceedings{jot1991digital,
    author = {J. M. Jot and A. Chaigne},
    booktitle = {Proc. Audio Eng. Soc. Conv.},
    title = {Digital delay networks for designing artificial reverberators},
    year = {1991},
    month = {Oct.},
    number = {3030}
}

@inproceedings{jot1992analysis,
    author = {J. M. Jot},
    booktitle = {Proc. IEEE Int. Conf. Acoust., Speech, Signal Process. (ICASSP)},
    title = {An analysis/synthesis approach to real-time artificial reverberation},
    year = {1992},
    month = {Mar.},
    volume = {2},
    pages = {221--224}
}

@article{gerzon1976unitary,
    author = {M. A. Gerzon},
    journal = {Electron. Lett.},
    title = {Unitary (energy-preserving) multichannel networks with feedback},
    year = {1976},
    month = {May},
    volume = {12},
    number = {11},
    pages = {278--279}
}

@article{schlecht2016lossless,
    author = {S. J. Schlecht and E. A. P. Habets},
    journal = {IEEE Trans. Signal Process.},
    title = {On lossless feedback delay networks},
    year = {2017},
    month = {Mar.},
    volume = {65},
    number = {6},
    pages = {1554--1564}
}

@article{schlecht2016feedback,
    author = {S. J. Schlecht and E. A. P. Habets},
    journal = {IEEE/ACM Trans. Audio, Speech, Lang. Process.},
    title = {Feedback delay networks: Echo density and mixing time},
    year = {2017},
    month = {Feb.},
    volume = {25},
    number = {2},
    pages = {280--289}
}

@article{schlecht2019modal,
    author = {S. J. Schlecht and E. A. P. Habets},
    journal = {IEEE Trans. Signal Process.},
    title = {Modal decomposition of feedback delay networks},
    year = {2019},
    month = {Oct.},
    volume = {67},
    number = {20},
    pages = {5340--5351}
}

@article{schlecht2020scattering,
    author = {S. J. Schlecht and E. A. P. Habets},
    journal = {IEEE/ACM Trans. Audio, Speech, Lang. Process.},
    title = {Scattering in feedback delay networks},
    year = {2020},
    month = {Apr.},
    volume = {28},
    pages = {1915--1924}
}

@inproceedings{prawda2019improved,
    author = {K. Prawda and V. Välimäki and S. Schlecht},
    booktitle = {Proc. Int. Conf. Digital Audio Effects (DAFx)},
    title = {Improved reverberation time control for feedback delay networks},
    year = {2019},
    month = {Sep.},
    pages = {1--8}
}

@article{das2021grouped,
    author = {O. Das and J. S. Abel},
    journal = {J. Audio Eng. Soc.},
    title = {Grouped feedback delay networks for modeling of coupled spaces},
    year = {2021},
    month = {Jul./Aug.},
    volume = {69},
    number = {7/8},
    pages = {486--496}
}

@article{das2023grouped,
    author = {O. Das and S. J. Schlecht and E. De Sena},
    journal = {IEEE/ACM Trans. Audio, Speech, Lang. Process.},
    title = {Grouped feedback delay networks with frequency-dependent coupling},
    year = {2023},
    month = {May},
    volume = {31},
    pages = {2004--2015}
}

@inproceedings{das2020delay,
    author = {O. Das and J. S. Abel and E. K. Canfield-Dafilou},
    booktitle = {Proc. Int. Conf. Digital Audio Effects (DAFx)},
    title = {Delay network architectures for room and coupled space modeling},
    year = {2020},
    month = {Sep.},
    pages = {234--241}
}

@inproceedings{das2026differentiable, 
title={Differentiable Grouped Feedback Delay Networks for Learning Direction and Position-Depedent Late Reverberation}, 
author={Das, Orchisama and Schlecht, Sebastian J and {Dal Santo}, Gloria and Cvetkovic, Zoran}, 
booktitle = {Proc. IEEE Int. Conf. Acoust., Speech, Signal Process. (ICASSP)},
year={2026},
note={Accepted}
}

@inproceedings{schlecht2017accurate,
    author = {S. J. Schlecht and E. A. P. Habets},
    booktitle = {Proc. Int. Conf. Digital Audio Effects (DAFx)},
    title = {Accurate reverberation time control in feedback delay networks},
    year = {2017},
    month = {Sep.},
    pages = {337--344}
}

@article{valimaki2016all,
    author = {V. Välimäki and J. D. Reiss},
    journal = {Appl. Sci.},
    title = {All about audio equalization: Solutions and frontiers},
    year = {2016},
    month = {May},
    volume = {6},
    number = {5},
    pages = {129}
}

@inproceedings{heldmann2021role,
    author = {J. Heldmann and S. J. Schlecht},
    booktitle = {Proc. Int. Conf. Digital Audio Effects (DAFx)},
    title = {The role of modal excitation in colorless reverberation},
    year = {2021},
    month = {Sep.},
    pages = {206--213}
}

@article{de2015efficient,
    author = {E. De Sena and H. Hacıhabiboğlu and Z. Cvetković and J. O. Smith},
    journal = {IEEE/ACM Trans. Audio, Speech, Lang. Process.},
    title = {Efficient synthesis of room acoustics via scattering delay networks},
    year = {2015},
    month = {Sep.},
    volume = {23},
    number = {9},
    pages = {1478--1492}
}

@article{atalay2022scattering,
    author = {T. B. Atalay and Z. S. Gül and E. De Sena and Z. Cvetković and H. Hacıhabiboğlu},
    journal = {IEEE/ACM Trans. Audio, Speech, Lang. Process.},
    title = {Scattering delay network simulator of coupled volume acoustics},
    year = {2022},
    month = {Feb.},
    volume = {30},
    pages = {582--593}
}

@article{bai2015late,
    author = {H. Bai and G. Richard and L. Daudet},
    journal = {IEEE/ACM Trans. Audio, Speech, Lang. Process.},
    title = {Late reverberation synthesis: From radiance transfer to feedback delay networks},
    year = {2015},
    month = {Dec.},
    volume = {23},
    number = {12},
    pages = {2260--2271}
}

@inproceedings{Das_Canfield-Dafilou_Abel_2019,
    author = {O. Das and E. K. Canfield-Dafilou and J. S. Abel},
    booktitle = {Proc. IEEE Workshop Appl. Signal Process. Audio Acoust. (WASPAA)},
    title = {On the behavior of delay network reverberator modes},
    year = {2019},
    month = {Oct.},
    pages = {50--54}
}

@inproceedings{dal2023differentiable,
    author = {G. {Dal Santo} and K. Prawda and S. Schlecht and V. Välimäki},
    booktitle = {Proc. Int. Conf. Digital Audio Effects (DAFx)},
    title = {Differentiable feedback delay network for colorless reverberation},
    year = {2023},
    month = {Sep.},
    pages = {244--251}
}

@article{santo2024feedback,
    author = {G. {Dal Santo} and K. Prawda and S. J. Schlecht and V. Välimäki},
    journal = {EURASIP J. Audio, Speech, Music Process.},
    title = {Optimizing tiny colorless feedback delay networks},
    year = {2024},
    month = {May},
    volume = {2025},
    number = {1},
    pages = {13}
}

@inproceedings{santo2024optimisation,
    author = {G. {Dal Santo} and B. Alary and K. Prawda and S. J. Schlecht and V. Välimäki},
    booktitle = {Proc. Int. Conf. Digital Audio Effects (DAFx)},
    title = {{RIR2FDN}: An improved impulse response analysis and synthesis},
    year = {2024},
    month = {Sep.},
    pages = {230--237}
}

@inproceedings{giampiccolo2024differentiable,
    author = {R. Giampiccolo and A. I. Mezza and A. Bernardini},
    booktitle = {Proc. Int. Conf. Digital Audio Effects (DAFx)},
    title = {Differentiable {MIMO} feedback delay networks for multichannel room impulse response modeling},
    year = {2024},
    month = {Sep.},
    pages = {278--285}
}

@article{mezza2024data,
    author = {A. I. Mezza and R. Giampiccolo and E. De Sena and A. Bernardini},
    journal = {EURASIP J. Audio Speech Music Process.},
    title = {Data-driven room acoustic modeling via differentiable feedback delay networks with learnable delay lines},
    year = {2024},
    month = {Oct.},
    volume = {2024},
    number = {1},
    pages = {1--20}
}

@inproceedings{mezza2025differentiable,
  title={Differentiable Scattering Delay Networks for Artificial Reverberation},
  author={A. I. Mezza and R. Giampiccolo and E. De Sena and A. Bernardini},
  booktitle={Proc. Int. Conf. Digital Audio Effects (DAFx)},
  pages={202--207},
  year={2025}
}

@article{lee2022differentiable,
    author = {S. Lee and H.-S. Choi and K. Lee},
    journal = {IEEE/ACM Trans. Audio, Speech, Lang. Process.},
    title = {Differentiable artificial reverberation},
    year = {2022},
    month = {Jul.},
    volume = {30},
    pages = {2541--2556}
}

@inproceedings{Mezza_Giampiccolo_Bernardini_2024,
    author = {A. I. Mezza and R. Giampiccolo and A. Bernardini},
    booktitle = {Proc. Int. Conf. Digital Audio Effects (DAFx)},
    title = {Modeling the frequency-dependent sound energy decay of acoustic environments with differentiable feedback delay networks},
    year = {2024},
    month = {Sep.},
    pages = {238--245}
}

@article{antoni2010orthogonal,
    author = {J. Antoni},
    journal = {J. Acoust. Soc. Amer.},
    title = {Orthogonal-like fractional-octave-band filters},
    year = {2010},
    month = {Feb.},
    volume = {127},
    number = {2},
    pages = {884--895}
}

@article{harma2000frequency,
    author = {A. Härmä and M. Karjalainen and L. Savioja and V. Välimäki and U. K. Laine and J. Huopaniemi},
    journal = {J. Audio Eng. Soc.},
    title = {Frequency-warped signal processing for audio applications},
    year = {2000},
    month = {Nov.},
    volume = {48},
    number = {11},
    pages = {1011--1031}
}

@article{Götz_Falcón_Pérez_Schlecht_Pulkki_2022,
    author = {G. Götz and R. Falcón Pérez and S. J. Schlecht and V. Pulkki},
    journal = {J. Acoust. Soc. Amer.},
    title = {Neural network for multi-exponential sound energy decay analysis},
    year = {2022},
    month = {Aug.},
    volume = {152},
    number = {2},
    pages = {942--953}
}

@inproceedings{Gotz_Kerimovs_Schlecht_Pulkki,
    author = {G. Götz and T. Kerimovs and S. J. Schlecht and V. Pulkki},
    booktitle = {Proc. AES Int. Conf. Audio Games},
    title = {Dynamic late reverberation rendering using the common-slope model},
    year = {2024},
    month = {Feb.}
}

@article{Gotz_Schlecht_Pulkki_2023,
    author = {G. Götz and S. J. Schlecht and V. Pulkki},
    journal = {IEEE/ACM Trans. Audio, Speech, Lang. Process.},
    title = {Common-slope modeling of late reverberation},
    year = {2023},
    month = {Sep.},
    volume = {31},
    pages = {3945--3957}
}

@article{scerbo2025modeling,
  author={Scerbo, M. and Schlecht, S.J. and Ali, R. and Savioja, L. and De Sena, E.},
  journal={IEEE Trans. Audio, Speech, Lang. Process.}, 
  title={Modeling Nonuniform Energy Decay Through the Modal Decomposition of Acoustic Radiance Transfer (MoD-ART)}, 
  year={2025},
  volume={33},
  number={},
  pages={3363-3376},
  doi={10.1109/TASLPRO.2025.3592322}}

@article{scerbo2025efficient,
  author={Scerbo, Matteo and Schlecht, Sebastian J. and Ali, Randall and Savioja, Lauri and De Sena, Enzo},
  journal={IEEE Trans. Audio, Speech Lang. Process.}, 
  title={Efficient Multichannel Auralization Based on the Modal Decomposition of Acoustic Radiance Transfer}, 
  year={2025},
  volume={33},
  number={},
  pages={4748-4759},
  doi={10.1109/TASLPRO.2025.3629242}}

@article{kirsch2023computationally,
  title={Computationally-efficient simulation of late reverberation for inhomogeneous boundary conditions and coupled rooms},
  author={C. Kirsch and T. Wendt and S. Van De Par and H. Hu and S. D. Ewert},
  journal={J. Audio Eng. Soc.},
  volume={71},
  number={4},
  pages={186--201},
  year={2023},
  publisher={Audio Engineering Society}
}

@inproceedings{Das_Cvetkovic_2025,
    author = {O. Das and G. {Dal Santo} and S. J. Schlecht and Z. Cvetković},
    title = {Neural network based interpolation of late reverberation in coupled spaces using the common slopes model},
    booktitle = {Proc. IEEE Workshop Appl. Signal Process. Audio Acoust. (WASPAA)},
    year = {2025},
    month = {October},
}

@article{alary2021perceptual,
    author = {B. Alary and P. Massé and S. J. Schlecht and M. Noisternig and V. Välimäki},
    journal = {J. Acoust. Soc. Amer.},
    title = {Perceptual analysis of directional late reverberation},
    year = {2021},
    month = {May},
    volume = {149},
    number = {5},
    pages = {3189--3199}
}

@article{xiang2009investigation,
    author = {N. Xiang and Y. Jing and A. C. Bockman},
    journal = {J. Acoust. Soc. Amer.},
    title = {Investigation of acoustically coupled enclosures using a diffusion-equation model},
    year = {2009},
    month = {Sep.},
    volume = {126},
    number = {3},
    pages = {1187--1198}
}

@article{hodgson1996diffuse,
    author = {M. Hodgson},
    journal = {Appl. Acoust.},
    title = {When is diffuse-field theory applicable?},
    year = {1996},
    month = {Oct.},
    volume = {49},
    number = {3},
    pages = {197--207}
}

@book{kuttruff2009roomacou,
    author = {H. Kuttruff},
    title = {Room Acoustics},
    edition = {5th},
    year = {2009},
    publisher = {CRC Press},
    address = {Abingdon, UK}
}

@article{meissner2009computer,
    author = {M. Meissner},
    journal = {Arch. Acoust.},
    title = {Computer modelling of coupled spaces: variations of eigenmodes frequency due to a change in coupling area},
    year = {2009},
    month = {Jun.},
    volume = {34},
    number = {2},
    pages = {157--168}
}

@article{neidhardt2022perceptual,
    author = {A. Neidhardt and C. Schneiderwind and F. Klein},
    journal = {Trends Hear.},
    title = {Perceptual matching of room acoustics for auditory augmented reality in small rooms---{L}iterature review and theoretical framework},
    year = {2022},
    month = {Jul.},
    volume = {26}
}

@article{Potter_Cvetkovic_DeSena_2022,
    author = {T. Potter and Z. Cvetković and E. De Sena},
    journal = {Front. Signal Process.},
    title = {On the relative importance of visual and spatial audio rendering on {VR} immersion},
    year = {2022},
    month = {Sep.},
    volume = {2},
    pages = {904866}
}

@inproceedings{schneiderwind2023effects,
    author = {C. Schneiderwind and M. Richter and N. Merten and A. Neidhardt},
    booktitle = {Proc. Immersive 3D Audio (I3DA)},
    title = {Effects of modified late reverberation on audio-visual plausibility and externalization in {AR}},
    year = {2023},
    month = {Sep.},
    pages = {1--9}
}

@inproceedings{abel2006simple,
    author = {J. S. Abel and P. Huang},
    booktitle = {Proc. Audio Eng. Soc. Conv.},
    title = {A simple, robust measure of reverberation echo density},
    year = {2006},
    month = {Oct.}
}

@inproceedings{abel2014modal,
    author = {J. S. Abel and S. Coffin and K. Spratt},
    booktitle = {Proc. Audio Eng. Soc. Conv.},
    title = {A modal architecture for artificial reverberation with application to room acoustics modeling},
    year = {2014},
    month = {Oct.}
}

@inproceedings{karjalainen2001more,
    author = {M. Karjalainen and H. Järveläinen},
    booktitle = {Proc. Audio Eng. Soc. Conv.},
    title = {More about this reverberation science: Perceptually good late reverberation},
    year = {2001},
    month = {Nov.}
}

@inproceedings{luo2022learning,
    author = {A. Luo and Y. Du and M. Tarr and J. Tenenbaum and A. Torralba and C. Gan},
    booktitle = {Adv. Neural Inf. Process. Syst.},
    title = {Learning neural acoustic fields},
    year = {2022},
    month = {Nov.},
    volume = {35},
    pages = {3165--3177}
}

@inproceedings{chen2023novel,
  title={Novel-view acoustic synthesis},
  author={C. Chen and A. Richard amd R. Shapovalov and V. K. Ithapu and N. Neverova and K. Grauman and A. Vedaldi},
  booktitle={Proc. IEEE/CVF Conf. Comp. Vis. Pat. Recog.},
  pages={6409--6419},
  year={2023}
}

@inproceedings{ick2025direction,
  title={Direction-Aware Neural Acoustic Fields for Few-Shot Interpolation of Ambisonic Impulse Responses},
  author={C. Ick and G. Wichern and Y. Masuyama and F. Germain and J. L. Roux},
  booktitle={Proc. Interspeech},
  year={2025}
}

@article{majumder2022few,
  title={Few-shot audio-visual learning of environment acoustics},
  author={Majumder, Sagnik and Chen, Changan and Al-Halah, Ziad and Grauman, Kristen},
  journal={Advances in Neural Information Processing Systems},
  volume={35},
  pages={2522--2536},
  year={2022}
}

@article{su2022inras,
  title={{INRAS}: Implicit neural representation for audio scenes},
  author={K. Su and M. Chen and E. Shlizerman},
  journal={Adv. Neural Inf. Process. Syst.},
  volume={35},
  pages={8144--8158},
  year={2022}
}

@inproceedings{richard2022deep,
    author = {A. Richard and P. Dodds and V. K. Ithapu},
    booktitle = {Proc. IEEE Int. Conf. Acoust., Speech, Signal Process. (ICASSP)},
    title = {Deep impulse responses: Estimating and parameterizing filters with deep networks},
    year = {2022},
    month = {May},
    pages = {3209--3213}
}

@inproceedings{kuznetsov2020differentiable,
    author = {B. Kuznetsov and J. D. Parker and F. Esqueda},
    booktitle = {Proc. Int. Conf. Digital Audio Effects (DAFx)},
    title = {Differentiable {IIR} filters for machine learning applications},
    year = {2020},
    month = {Sep.},
    pages = {297--303}
}

@misc{lei2016layer,
    author = {J. Lei Ba and J. R. Kiros and G. E. Hinton},
    title = {Layer normalization},
    howpublished = {arXiv preprint arXiv:1607.06450},
    year = {2016},
    month = {Jul.}
}

@inproceedings{he2015delving,
    author = {K. He and X. Zhang and S. Ren and J. Sun},
    booktitle = {Proc. IEEE Int. Conf. Comput. Vis. (ICCV)},
    title = {Delving deep into rectifiers: Surpassing human-level performance on {ImageNet} classification},
    year = {2015},
    month = {Dec.},
    pages = {1026--1034}
}

@inproceedings{colonel2022direct,
    author = {J. T. Colonel and C. J. Steinmetz and M. Michelen and J. D. Reiss},
    booktitle = {Proc. IEEE Int. Conf. Acoust., Speech, Signal Process. (ICASSP)},
    title = {Direct design of biquad filter cascades with deep learning by sampling random polynomials},
    year = {2022},
    month = {May},
    pages = {3104--3108}
}

@article{hayes2024review,
    author = {B. Hayes and J. Shier and G. Fazekas and A. McPherson and C. Saitis},
    journal = {Front. Signal Process.},
    title = {A review of differentiable digital signal processing for music and speech synthesis},
    year = {2024},
    month = {Feb.},
    volume = {3},
    pages = {1284100}
}

@article{yang2022audio,
    author = {J. Yang and A. Barde and M. Billinghurst},
    journal = {J. Audio Eng. Soc.},
    title = {Audio augmented reality: A systematic review of technologies, applications, and future research directions},
    year = {2022},
    month = {Oct.},
    volume = {70},
    number = {10},
    pages = {788--809}
}
%%%%%%%%%%%%%%%%%%%%%%%%%%%%%%%%%%%%%%%%%%%%%%%%
% \IEEEtriggeratref{999}
% \input{sections/biography}

\end{document}